%%%%%%%%% TEX-file

\input amstex
\magnification\magstep1
\parskip=6pt
\parindent=0pt
\tolerance16000
\loadmsam
\loadmsbm
\input amssym.def
\input amssym.tex

\UseAMSsymbols
\loadbold

\font\sans=cmss10

\def\z{\zeta}
\def\pagenumber{\footline={\hss\tenrm\folio\hss}}

\def\half{{1\over 2}}
\def\third{{1\over 3}}

\def\A{{\Cal A}}
\def\B{{\Cal B}}
\def\C{{\Cal C}}
\def\D{{\Cal D}}
\def\H{{\Cal H}}
\def\M{{\Cal M}}
\def\O{{\Cal O}}
\def\Q{{\Cal Q}}
\def\R{{\Cal R}}
\def\U{{\Cal U}}
\def\V{{\Cal V}}

\def\Op{{{\Cal O}^\prime}}
\def\Dp{{{\Cal D}^\prime}}
\def\Cp{{{\Cal C}^\prime}}
\def\rs{{\Bbb P}}

\def\tV{\widetilde\V}

\def\sV{{\hbox{\sans V}}}
\def\sH{{\hbox{\sans H}}}

\def\ref#1{$^{[#1]}$}

\def\half{{1\over 2}}
\def\third{{1\over 3}}

\def\ie{{\it i.e.}}

\def\Cop{{\Bbb C}}
\def\Nop{{\Bbb N}}
\def\Pop{{\Bbb P}}
\def\Zop{{\Bbb Z}}

\def\Hom{\hbox{Hom}}
\def\End{\hbox{End}}

\def\enbolden#1{{\boldsymbol#1}}
\def\bpsi{\enbolden\psi}
\def\bPsi{\enbolden\Psi}
\def\bphi{\enbolden\phi}
\def\bchi{\enbolden\chi}
\def\balpha{\enbolden\alpha}
\def\bX{\enbolden X}
\def\bz{\enbolden z}

\def\empty{\varnothing}

\nopagenumbers
\rightline{DAMTP-1998-135}
\rightline{hep-th/9810019}
\vskip 80pt
\centerline {\bf AXIOMATIC CONFORMAL FIELD THEORY}
\vskip 45 pt
\centerline {\bf Matthias R.\ Gaberdiel and Peter Goddard}
\vskip 12 pt
\centerline {\bf Department of Applied Mathematics and Theoretical
Physics} 
\centerline {\bf University of Cambridge, Silver Street}
\centerline {\bf Cambridge, CB3 9EW, U.K.}
\vskip 50 pt
\centerline {\bf ABSTRACT}
\vskip 6 pt
{\rightskip=10 true mm \leftskip=10 true mm  \noindent

A new rigorous approach to conformal field theory is presented.
The basic objects are families of complex-valued amplitudes,
which define a meromorphic conformal field theory (or chiral
algebra)  and which lead naturally to the definition of 
topological vector spaces, between which vertex operators act
as continuous operators. In fact, in order to develop the
theory, M\"obius invariance rather than full conformal
invariance is required but it is shown that every M\"obius
theory can be extended to a conformal theory by the
construction of a Virasoro field.

In this approach, a representation of a conformal field theory is 
naturally defined in terms of a family of amplitudes with appropriate
analytic properties. It is shown that these amplitudes can also be
derived from a suitable collection of states in the meromorphic
theory. Zhu's algebra then appears naturally as the algebra of
conditions which states defining highest weight representations must
satisfy. The relationship of the representations of Zhu's algebra to
the classification of highest weight representations is explained.

}
\vskip 30 pt
\centerline{October  1998}
\vfill\eject
\pagenumber

\leftline {\bf 1. Introduction}
\vskip6pt
\nobreak

Conformal field theory has been a subject which has attracted a great
deal of attention in the last thirty years, much of the interest being
motivated by its importance in string theory. Its role in string
theory goes back to its very beginning when Veneziano\ref{1} proposed
a form for the scattering amplitude for four particles, quickly
generalised to $n$-particle amplitudes, which could conveniently be
expressed as integrals in the complex plane of meromorphic
functions\ref{2}. From these amplitudes, the space of states was
obtained by factorisation.

The power of two-dimensional conformal field theory was conclusively
demonstrated by the work of Belavin, Polyakov and
Zamolodchikov\ref{3}. They set a general framework for its study
which was further developed by Moore and Seiberg\ref{4}, in
particular. This approach is couched within the general language of
quantum field theory. Not least because it is possible to establish
very strong results in conformal field theory, it is very desirable to
have a precise mathematical context within which they can be
established. Rigorous approaches to conformal field theory have been 
developed, broadly speaking, from three different stand points: a
geometrical approach initiated by Segal\ref{5}; an algebraic approach
due to Borcherds\ref{6,7}, Frenkel, Lepowsky and Meurman\ref{8}
and developed further by Frenkel, Huang and Lepowsky\ref{9} and
Kac\ref{10}; and an functional analytic approach in which techniques
from algebraic quantum field theory are employed and which has been
pioneered by Wassermann\ref{11} and Gabbiani and Fr\"ohlich\ref{12}.  

Each of these three approaches produces a different perspective on
conformal field and each facilitates the appreciation of its deep
connections with other parts of mathematics, different in the three
cases. Here we present a rigorous approach closely related to the way
conformal field theory arose at the birth of string theory. It is a
development of earlier studies of meromorphic conformal field
theory\ref{13}. Starting from a family of `amplitudes', which are
functions of $n$ complex variables and describe the vacuum
expectation values of $n$ fields associated with certain basic states,
the full space of states of the theory is obtained by factorising
these amplitudes in a certain sense. 

The process of reconstructing the space of states from the vacuum
expectation values of fields is familiar from axiomatic quantum
field theory. In the usual Osterwalder-Schrader framework of Euclidean
quantum field theory\ref{14}, the reflection positivity axiom
guarantees that the resulting space of states has the structure of a
Hilbert space. (In the context of conformal field theory this approach
has been developed by Felder, Fr\"ohlich and Keller\ref{15}.) 
In the present approach, the construction of the space of states
depends only on the meromorphicity of the given family of amplitudes,
$\A$, and positivity is not required for the basic development of the
theory. 

The spaces of states that are naturally defined are not Hilbert spaces
but topological vector spaces, their topology being determined by 
requirements designed to ensure meromorphic amplitudes. (Recently
Huang has also introduced topological vector spaces, which are related
to ours, but from a different point of view\ref{16}.)   They are also such
that one can introduce fields (`vertex operators') for the basic states
which are continuous operators. The locality property of these vertex
operators is a direct consequence of the locality assumption about the
family of amplitudes, $\A$, and this is then sufficient to prove the
duality property (or Jacobi identity) of the vertex operators\ref{13}.

To develop the theory further, we need to assume that the basic
amplitudes, $\A$, are M\"obius invariant. This enables us to define
vertex operators for more general states, modes of vertex operators
and a Fock space which contains the essential algebraic content of the
theory. This Fock space also enables us to define the concept of the
equivalence of conformal field theories. The assumptions made so far
are very general but if we assume that the amplitudes satisfy a
cluster decomposition property we place much more severe restrictions
on the theory, enabling us, in particular, to prove the uniqueness of
the vacuum state.  

Nothing assumed so far implies that the theory has a conformal
structure, only one of M\"obius symmetry. However, we show that it is
always possible to extend the theory in such a way that it acquires a
conformal structure. (For theories with a conformal structure this
leaves the theory unchanged.) A conformal structure is necessary if we
want to be able to define the theory on higher genus Riemann surfaces
(although this is not discussed in the present paper). For this
purpose, we also need to introduce the concept of a representations of
a conformal (or rather a M\"obius) field theory. Developing an idea of
Montague, we show that any representation corresponds to a state in
the space of states of the theory\ref{17}. This naturally poses the
question of what conditions a state has to satisfy in order to define
a representation. For the case of highest weight representations, the
conditions define an associative algebra which is that originally
introduced by Zhu\ref{18}. It is the main content of Zhu's Theorem that
this algebra can be defined in terms of the algebraic Fock space.

The plan of the paper is as follows. In Section 2, we introduce the
basic assumptions about the family of amplitudes, $\A$, and construct
the topological vector space of states and the vertex operators for
the basic states. In Section 3, M\"obius invariance and its
consequences are discussed. In Section 4, we define modes of vertex
operators and use them to construct Fock spaces and thus to define the
equivalence of theories. Examples of conformal field theories are
provided in Section 5: the $U(1)$ theory; affine Lie algebra theory;
the Virasoro theory; lattice theories; and an example which does not
have a conformal structure. In Section 6, the assumption of cluster
decomposition is introduced and in Section 7 we show how to extend a
M\"obius invariant theory to make it conformally invariant. In Section
8, we define what is meant by a representation and show how any
representation can be characterised by a state in the theory. In
Section 9, we define the idea of a M\"obius covariant representation
and the notion of equivalence for representations. An example of a
representation is given in Section 10. In Section 11, we introduce
Zhu's algebra and explain the significance of Zhu's Theorem in our
context. Further developments, which are to be the subject of a future
paper\ref{19}, are surveyed in Section 12. There are seven appendices
in which some of the more technical details are described.

\vskip24pt
\leftline{\bf 2. Amplitudes, Spaces and Vertex Operators}
\vskip6pt

The starting point for our approach is a collection of functions,
which are eventually to be regarded as the vacuum expectation values
of the fields associated with a certain basic set of states which
generate the whole theory. We shall denote the space spanned by such
states by $V$. In terms of the usual concepts of conformal field
theory, $V$ would be a subspace of the space of quasi-primary
states. $V$ can typically be taken to be finite-dimensional but this
is not essential in what follows.  (If it is infinite-dimensional,
we shall at least assume that the algebraic dimension of $V$ is
countable, that is that the elements of $V$ consist of finite
linear combinations of a countable basis.)

We suppose that $V$ can be regarded as the direct sum of a collection
of subspaces, $V_h$, to each of which we can attach an integer,
$h$, called  the conformal weight of the states in that subspace,
so that $V=\bigoplus_h V_h$. This is equivalent to saying that 
we have a diagonalisable operator $\delta:V\rightarrow V$, with 
eigenspaces $V_h=\{\psi\in V: \delta\psi=h\psi\}$.

We also suppose that for any positive integer $n$, and any finite
collection of vectors $\psi_i\in V_{h_i}$, and $z_i \in \rs$ (the
Riemann Sphere),where $i=1, \ldots , n$, we have a density 
$$f(\psi_1,\ldots,\psi_n;z_1,\ldots,z_n)
\equiv\langle V(\psi_1,z_1)V(\psi_2,z_2)\cdots V(\psi_n,z_n)\rangle
\prod_{j=1}^n\left(dz_j\right)^{h_j}\,.
\eqno(1)$$
Here  $\langle V(\psi_1,z_1)V(\psi_2,z_2)\cdots V(\psi_n,z_n)\rangle$
is merely a suggestive notation for what will in the end acquire an
interpretation as the vacuum expectation value of a product of
fields. These ``amplitudes'' are assumed to be multilinear in 
$\psi_i$, invariant under the exchange of $(\psi_i,z_i)$ with
$(\psi_j,z_j)$, and analytic in $z_i$, save only for possible
singularities occurring at $z_i=z_j$ for $i\neq j$, which we shall
assume to be poles of finite order (although one could consider
generalisations in which the amplitudes are allowed to have essential
singularities). Because of the independence of order of the
$(\psi_j,z_j)$, we can use the notation  
$$
f(\psi_1,\ldots,\psi_n;z_1,\ldots,z_n)
=\left\langle \prod_{j=1}^nV(\psi_j,z_j)\right\rangle
\prod_{j=1}^n\left(dz_j\right)^{h_j}\,.\eqno(2)
$$
We denote the collection of these densities, and the theory we develop
from them, by $\A=\{f\}$. We may assume that if all amplitudes in $\A$
involving a given $\psi\in V$ vanish then $\psi=0$ (for, if this is
not so, we may replace $V$ by its quotient by the space of all vectors
$\psi\in V$ which are such that all amplitudes involving $\psi$
vanish). 

We use these amplitudes to define spaces of states associated with
certain subsets $\C$ of the Riemann Sphere $\rs$. We can picture these
spaces as consisting of states generated by fields acting at points of
$\C$. First introduce the set, $\B_\C$, whose elements are labelled by
finite collections of $\psi_i\in V_{h_i}$, $z_i\in\C\subset \rs$, 
$i =1,\ldots, n$, $n\in\Nop$ and $z_i\ne z_j$ if $i\ne j$;  we denote
a typical element $\bpsi\in\B_\C$ by 
$$
\bpsi = V(\psi_1,z_1)V(\psi_2,z_2)\cdots V(\psi_n,z_n)\Omega
\equiv \prod_{i=1}^n V(\psi_i,z_i)\Omega\,.\eqno(3)
$$
We shall immediately identify $\bpsi\in\B_\C$ with the other elements
of $\B_\C$ obtained by replacing each $\psi_j$ in (3) by
$\mu_j\psi_j$, $1\leq j\leq n$, where $\mu_j\in\Cop$ and
$\prod_{j=1}^n\mu_j=1$. 

Next we introduce the free (complex) vector space on $\B_\C$, 
\ie\ the complex vector space with {\it basis} $\B_\C$ that is  
consisting of formal finite linear combinations 
$\Psi =\sum_j\lambda_j\bpsi_j$, $\lambda_j\in\Cop$, $\bpsi_j\in\B_\C$;
we denote this space by  $\V_\C$.  

The vector space $\V_\C$ is enormous, and, intuitively, as we consider
more and more complex combinations of the basis vectors, $\B_\C$, we  
generate vectors which are very close to one another. To measure this
closeness, we need in essence to use suitably chosen amplitudes as
test functions. To select a collection of linear functionals which we
may use to construct from $\V_\C$ a space in which we have some
suitable  idea of topology, we select another subset $\O\subset\rs$
with $\O\cap\C=\empty$, where $\O$ is open, and we suppose further
that the interior of $\C$, $\C^o$, is not empty. Let 
$$\bphi = V(\phi_1,\zeta_1)V(\phi_2,\zeta_2)\cdots
V(\phi_m,\zeta_m)\Omega \in \B_\O \,,\eqno(4)$$
where $\phi_j\in V_{k_j}$, $j=1,\ldots, m$. Each $\bphi\in\B_\O$
defines a map on $\bpsi\in\B_\C$ by 
$$\eta_{\bphi}(\bpsi)=(\bphi,\bpsi)
=\left\langle \prod_{i=1}^mV(\phi_i,\zeta_i)
\prod_{j=1}^nV(\psi_j,z_j)\right\rangle\,,\eqno(5)$$
which we can use as a contribution to our measure of nearness of
vectors in $\V_\C$. [Strictly speaking, this map defines a density
rather than a function, so that we should  really be considering 
$\eta_{\bphi}(\bpsi) \prod_{i=1}^m\left(d\zeta_i\right)^{k_i}
\prod_{j=1}^n\left(dz_j\right)^{h_j}\,.]$

For each $\bphi\in\B_\O$, $\eta_{\bphi}$ extends by linearity to a map
$\V_\C\rightarrow\Cop$, provided that $\O\cap\C=\empty$. We use these
linear functionals to define our concept of closeness or, more
precisely, the topology of our space. To make sure that we end up with
a space which is complete, we need to consider sequences of elements
of $\V_\C$ which are convergent in a suitable sense. Let $\tV_\C$ be
the space of sequences $\bPsi=(\Psi_1,\Psi_2,\ldots)$,
$\Psi_j\in\V_\C$. We consider the subset $\tV_\C^\O$ of such sequences
$\bPsi$ for which $\eta_{\bphi}(\Psi_j)$ converges on subsets of
$\bphi$ of the form  
$$\{\bphi =V(\phi_1,\zeta_1)V(\phi_2,\zeta_2)\cdots
V(\phi_m,\zeta_m)\Omega : \zeta_j\in K, |\zeta_i - \zeta_j| 
\geq \epsilon, i\ne j\},\eqno(6)$$ 
where for each collection of $\phi_j$, $\epsilon>0$ and a compact subset
$K\subset \O$, the convergence is uniform in the (compact) set
$$
\{(\zeta_1,\ldots,\zeta_m):\zeta_j\in K, |\zeta_i - \zeta_j| 
\geq \epsilon, i\ne j\}\,. \eqno(7)$$
If $\bPsi\in\tV_\C^\O$, the limit 
$$\lim_{j\rightarrow\infty}\eta_\bphi(\Psi_j)\eqno(8)$$ 
is necessarily an analytic function of the $\zeta_j$, for
$\zeta_j\in\O$, with singularities only at $\zeta_i =\zeta_j$, 
$i\ne j$. (Again these could in principle be essential singularities,
but the assumption of the cluster decompostion property, made in
Section 6, will imply that these are only poles of finite order.) We
denote this function by $\eta_\bphi(\bPsi)$. [A necessary and
sufficient condition for uniform convergence on the compact set (7) is
that the functions $\eta_{\bphi}(\Psi_j)$ should be both convergent in
the compact set and locally uniformly bounded, \ie\ each point of (7)
has a neighbourhood in which $\eta_{\bphi}(\Psi_j)$ is bounded
independently of $j$; see Appendix~A for further details.]    

It is natural that we should regard two such sequences 
$\bPsi^1 = (\Psi^1_i)$ and $\bPsi^2 = (\Psi^2_i)$ as equivalent if 
$$\lim_{j\rightarrow\infty}\eta_\bphi(\Psi^1_j) =
\lim_{j\rightarrow\infty}\eta_\bphi(\Psi^2_j)\,,\eqno(9)$$
\ie\ $\eta_\bphi(\bPsi^1)=\eta_\bphi(\bPsi^2)$, 
for each $\bphi\in\B_\O$. We identify such equivalent sequences, and
denote the space of them by $\V_\C^\O$. 

The space $\V_\C^\O$ has a natural topology: we define a sequence
$\chi_j\in\V_\C^\O$, $j = 1,2, \ldots $, to be convergent if, for each
$\bphi\in\B_\O$, $\eta_\bphi(\chi_j)$ converges uniformly on each
(compact) subset of the form (7). The limit  
$$\lim_{j\rightarrow\infty}\eta_\bphi(\chi_j)\eqno(10)$$
is again necessarily a meromorphic function of the $\zeta_j$, for
$\zeta_j\in\O$, with poles only at $\zeta_i =\zeta_j$, $i\ne j$. 
Provided that the limits of such sequences are always in $\V^\O_\C$,
\ie 
$$\lim_{j\rightarrow\infty}\eta_\bphi(\chi_j)
=\eta_\bphi(\chi),\qquad\hbox{for some }\chi\in\V_\C^\O\eqno(11)$$
we can define the topology by defining its closed subsets to be those
for which the limit of each convergent sequence of elements in the
subset is contained within it. In fact we do not have to incorporate
the need for the limit to be in $\V_\C^\O$, because it is so
necessarily; we show this in Appendix~B. [As we note in this Appendix,
this topology on $\V^\O_\C$ can be induced by a countable family of
seminorms of the form 
$||\chi||_n=\max_{1\leq i \leq n} \max_{\zeta_{i_j}} 
|\eta_{\bphi_i}(\chi)|$, where the $\phi_{i_j}$ in $\bphi_i$ are
chosen from finite subsets of a countable basis and the $\zeta_{i_j}$ are
in a compact set of the form (7).] 

$\B_\C$ can be identified with a subset of $\tV_\C^\O$ (using constant
sequences), and this has an image in $\V_\C^O$. It can be shown that
this image is necessarily faithful provided that we  assume the
cluster property introduced in Section 6. In any case, we shall assume
that this is the case in what follows and identify $\B_\C$ with its
image in $\V_\C^\O$. There is a common vector
$\Omega\in\B_\C\subset\V_\C^\O$ for all $\C,\O$ which is called 
the {\it vacuum vector}. The linear span of $\B_\C$ is dense in
$\V^\O_\C$, (\ie\ it is what is called a {\it total space}).  With
this identification, the image of $\B_\C$ in $\V^\O_\C$, $\bpsi$,
defined as in (3), depends linearly on the vectors $\psi_j\in V$.

A key result in our approach is that, for suitable $\O$, $\V_\C^\O$
does not depend on $\C$. This is an analogue of the Reeh-Schlieder
Theorem of Axiomatic Quantum Field Theory. In our context it is
basically a consequence of the fact that any meromorphic function is
determined by its values in an arbitrary open set. Precisely, we have
the result:  

{\bf Theorem 1:} $\V^\O_\C$ is independent of $\C$ if the complement of
$\O$ is path connected.

The proof is given in Appendix~C. In the following we shall mainly
consider the case where the complement of $\O$ is path-connected and,
in this case, we denote $\V^\O_\C$ by $\V^\O$.

The definition of $\eta_{\bphi}:\V_\C\rightarrow\Cop$, $\tV_\C$ and,
in particular, $\V_\C^\O$ all depend, at least superficially, on the
particular coordinate chosen on $\Pop$, that is the particular
identification of $\Pop$ with $\Cop\cup\{\infty\}$. However the
coefficients with which elements of $\B_\C$ are combined to constitute
elements of $\V_\C$ should be regarded as densities and then a change
of coordinate on $\Pop$ induces an endomorphism of $\V_\C$ which
relates the definitions of the space $\V^\O_\C$ which we would get
with the different choices of coordinates, because $\eta_{\bphi}$ only
changes by an overall factor (albeit a function of the $\zeta_i$). 
In this way $\V^\O_\C$, {\it etc.} can be regarded as coordinate
independent.

Suppose that $\O\subset\Op$ and $\C\cap\Op=\empty$ with
$\C^o\ne\empty$. Then if a sequence $\bPsi = (\Psi_j) \in \tV_\C$ is
such that $\eta_\bphi(\Psi_j)$ is convergent for all $\bphi\in\B_\Op$
it follows that it is convergent for all
$\bphi\in\B_\O\subset\B_\Op$. In these circumstances, if
$\eta_\bphi(\bPsi)$ vanishes for all $\bphi\in\B_\O$, it follows that
$\eta_{\bphi^\prime}(\bPsi)$ will vanish for all
$\bphi^\prime\in\B_\Op$, because each $\eta_{\bphi^\prime}(\bPsi)$ is
the analytic continuation of $\eta_\bphi(\bPsi)$, for some
$\bphi\in\B_\O$; the converse is also immediate because
$\B_\O\subset\B_\Op$. Thus members of an equivalence class in
$\V^\Op_\C$ are also in the same equivalence class in $\V^\O_\C$. We
thus have an injection $\V^\Op\rightarrow\V^\O$, and we can regard 
$\V^\Op\subset\V^\O$. Since $\B_\C$ is dense in $\V^\O$, it follows that
$\V^\Op$ is also.  

Given a subset $\C\subset\rs$ with $\C^o\ne\empty$, $\B_\C$
is dense in a collection of spaces $\V^\O$, with $\C\cap\O=\empty$.
Given open sets $\O_1$ and $\O_2$ such that the complement of
$\O_1\cup\O_2$ contains an open set, $\B_\C$ will be dense in both
$\V^{\O_1}$ and $\V^{\O_2}$ if $\C$ is contained in the complement
of $\O_1\cup\O_2$ and $\C^o\ne\empty$. The collection
of topological vector spaces $\V^\O$, where $\O$ is an open subset
of the Riemann sphere whose complement is path-connected, forms in
some sense the space of states of the meromorphic field
theory we are considering.

A {\it vertex operator} can be defined for $\psi\in V$ as an operator 
$V(\psi,z):\V^\O\rightarrow\V^{\Op}$, where $z\in\O$ but
$z\notin\Op\subset\O$, by defining its action on the dense subset
$\B_\C$, where $\C\cap\O=\empty$ 
$$V(\psi,z)\bpsi=V(\psi,z)V(\psi_1,z_1)V(\psi_2,z_2)\cdots
V(\psi_n,z_n)\Omega\eqno(12)$$
and $\bpsi\in\B_\C$. The image is in $\V_{\Cp}$ for any $\Cp\supset\C$
which contains $z$, and we can choose $\Cp$ such that
$\Cp\cap\Op=\empty$. This then extends by linearity to a map
$\V_\C\rightarrow\V_{\Cp}$. To show that it induces a map
$\V^\O_\C\rightarrow\V^{\Op}_{\Cp}$ we need to show that if 
$\Psi^j,\,\in\V^\O_\C,\,\rightarrow 0$ as $j\rightarrow\infty$, 
then $V(\psi,z)\Psi^j\rightarrow 0$ as $j\rightarrow\infty$; 
\ie\ if $\eta_\bphi(\Psi^j)\rightarrow 0$ for all $\bphi\in\B_\O$, then
$\eta_{\bphi^\prime}(V(\psi,z)\Psi^j)\rightarrow 0$ for all
$\bphi^\prime\in\B_{\Op}$. But 
$\eta_{\bphi^\prime}(V(\psi,z)\Psi^j)=\eta_{\bphi}(\Psi^j)$ where
$\bphi = V(\psi,z)\bphi^\prime \in \B_\O$ and so tends to zero as
required. It is straightforward to show that the vertex operator
$V(\psi,z)$ is continuous. We shall refer to these vertex operators
also as {\it meromorphic fields}. 

It follows directly from the invariance of the amplitudes under
permutations that

{\bf Proposition 2:} If $z, \zeta\in\O$, $z\ne \zeta$, and 
$\phi,\psi\in V$, then 
$$V(\phi,z)V(\psi,\zeta) = V(\psi,\zeta)V(\phi,z)\eqno(13)$$
as an identity on $\V^\O$. 

This result, that the vertex operators, $V(\psi,z)$, commute at
different $z$ in a (bosonic) meromorphic conformal field theory, is
one which should hold morally, but normally one has to attach a
meaning to it in some other sense, such as  analytic continuation
(compare for example [13]).

\vskip24pt
{\bf 3. M\"obius Invariance}
\vskip6pt

In order to proceed much further, without being dependent in some
essential way on how the Riemann sphere is identified with the complex
plane (and infinity), we shall need to assume that the amplitudes $\A$
have some sort of M\"obius invariance. We shall say that the densities
in $\A$ are invariant under the M\"obius transformation $\gamma$,
where  
$$\gamma(z)={az+b\over cz+d},\eqno(14)$$
(and we can take $ad-bc=1$), provided that the densities in (2) satisfy
$$\left\langle \prod_{j=1}^nV(\psi_j,z_j)\right\rangle
\prod_{j=1}^n(dz_j)^{h_j}=
\left\langle \prod_{j=1}^nV(\psi_j,\zeta_j)\right\rangle
\prod_{j=1}^n(d\zeta_j)^{h_j}, \quad\hbox{where } 
\zeta_j=\gamma(z_j)\,,
\eqno(15a)$$
\ie
$$\left\langle \prod_{j=1}^nV(\psi_j,z_j)\right\rangle=
\left\langle \prod_{j=1}^nV(\psi_j,\gamma(z_j))\right\rangle
\prod_{j=1}^n\left(\gamma'(z_j)\right)^{h_j}\,.\eqno(15b)$$
Here $\psi_j\in V_{h_j}$. The M\"obius transformations form the group
$\M\cong \hbox{SL(2,$\Cop$)}/\Zop_2$.

If $\A$ is invariant under the M\"obius transformation $\gamma$, 
we can define an operator $U(\gamma):\V^\O\rightarrow\V^{\O_\gamma}$,
where $\O_\gamma=\{\gamma(z) : z \in \O\}$, by defining it on the
dense subset $\B_\C$ for some $\C$ with $\C\cap\O=\empty$, by 
$$U(\gamma)\bpsi=
\prod_{j=1}^n V(\psi_j,\gamma(z_j))
\prod_{j=1}^n\left(\gamma'(z_j)\right)^{h_j}\Omega\,,\eqno(16)$$
where $\bpsi=V(\psi_1,z_1)\cdots V(\psi_n,z_n)\Omega\in\B_\C$. Again,
this extends by linearity to a map defined on $\V_\C$, and to show
that it defines a map $\V_\C^\O\rightarrow\V_{\C_\gamma}^{\O_\gamma}$,
where $\C_\gamma=\{\gamma(z) : z \in \C\}$, we again need to show
that if $\eta_\bphi(\Psi_j)\rightarrow 0$ for all $\bphi\in\B_\O$,
then $\eta_{\bphi^\prime}(U(\gamma)\Psi_j)\rightarrow 0$ for all 
$\bphi^\prime\in\B_{\O_\gamma}$. By the assumed invariance under
$\gamma$, we have 
$\eta_{\bphi^\prime}(U(\gamma)\Psi_j)=\eta_\bphi(\Psi_j)$, where  
$\bphi= U(\gamma^{-1})\bphi^\prime$, and the result follows.

It follows immediately from the definition of $U(\gamma)$ that
$U(\gamma)\Omega=\Omega$ (where we have identified $\Omega\in\V^\O$
with $\Omega\in \V^{\O\gamma}$ as explained in Section 2).
Furthermore, 
$$U(\gamma)V(\psi,z)U(\gamma^{-1})
=V(\psi,\gamma(z))\gamma'(z)^{h},\quad 
\hbox{for }\psi\in V_h.\eqno(17)$$

By choosing a point $z_0\notin\O$, we can identify $V$ with a
subspace of $\V^\O$ by the map $\psi\mapsto V(\psi,z_0)\Omega$;
this map is an injection provided that $\A$ is invariant under
an infinite subgroup of $\M$ which maps $z_0$ to an infinite
number of distinct image points. For, if  
$$\langle \prod_{i=1}^{n} V(\psi_i,z_i) V(\psi,\zeta) \rangle
\eqno(18) $$
vanishes for $\zeta=z_0$ for all $\psi_i$ and $z_i$, then
by the invariance property, the same holds for an infinite
number of $\zeta$'s. Regarded as a function of $\zeta$, (18)
defines a meromorphic function with infinitely many zeros;
it therefore vanishes identically, thus implying that $\psi=0$.

In the following we shall use elements of $\hbox{SL(2,$\Cop$)}$ to
denote the corresponding elements of $\M$ where no confusion will
result, so that 
$$\hbox{if}\qquad \gamma=\pmatrix a&b\\ c&d\endpmatrix, 
\qquad \gamma(z)={az+b\over cz+d}\,.\eqno(19)$$

An element of $\M$ has either one or two fixed points or is the
identity. The one-parameter complex subgroups of $\M$ are either
conjugate to the translation group $z\mapsto z+\lambda$ (one fixed
point) or  the dilatation group $z\mapsto e^\lambda z$ (two fixed
points). 

Now, first, consider a theory which is invariant under the translation
group $z\mapsto \tau_\lambda(z) = z+\lambda$. Then, if 
$\tau_\lambda = e^{\lambda L_{-1}}$, and we do not distinguish between
$U(L_{-1})$ and $L_{-1}$ in terms of notation, from (17) we have 
$$e^{\lambda L_{-1}} V(\psi,z)e^{-\lambda L_{-1}}=
V(\psi,z+\lambda)\,.\eqno(20)$$ 
[If, instead, we had a theory invariant under a subgroup of the
M\"obius group conjugate to the translation group,
$\{\gamma_0^{-1}\tau_\lambda\gamma_0:\lambda\in\Cop\}$ say, and if
$\zeta=\gamma_0(z)$, $\zeta\mapsto \zeta'=\zeta+\lambda$ under 
$\gamma_0^{-1}\tau_\lambda\gamma_0$; then, if
$\widehat{V}(\psi,\zeta) = V(\psi,z)\gamma_0'(z)^{-h}$ and 
$\hat L_{-1}= \gamma_0^{-1}L_{-1}\gamma_0$, then 
$e^{\lambda \hat L_{-1}} \widehat{V}(\psi,\zeta)e^{-\lambda \hat
L_{-1}} = \widehat{V}(\psi,\zeta+\lambda)$.]

Consider now a theory which is invariant under the whole M\"obius
group. We can pick a group conjugate to the translation group, and we
can change coordinates so that $z=\infty$ is the fixed point of the
selected translation group. (In particular, this defines an
identification of $\Pop$ with $\Cop\cup\{\infty\}$ up to a Euclidean
or scaling transformation of $\Cop$.)  If we select a point $z_0$ to
define the injection $V\rightarrow\V^\O$, $z_0\notin\O$, we have 
effectively selected two fixed points. Without loss of generality, we
can choose $z_0=0$. Then 
$$\psi=V(\psi,0)\Omega\in\V^\O\,.\eqno(21)$$

We can then introduce naturally two other one-parameter groups, one
generated by $L_0$ which fixes both $0$ and $\infty$ (the group of
dilatations or scaling transformations), and another which  fixes only
$0$, generated by $L_1$ (the group of special conformal
transformations). Then 
$$e^{\lambda L_{-1}}(z) = z+\lambda,\qquad
e^{\lambda L_0}(z)=e^\lambda z,\qquad
 e^{\lambda L_1}(z) = {z\over 1-\lambda z}\,,\eqno(22a)$$
$$e^{\lambda L_{-1}}=\pmatrix 1&\lambda\\ 0&1\endpmatrix,\qquad
e^{\lambda L_0}=\pmatrix
e^{\half\lambda}&0\cr0&e^{-\half\lambda}\endpmatrix, 
\qquad
e^{\lambda L_1} = \pmatrix 1&0\\ -\lambda&1\endpmatrix\,,\eqno(22b)$$
and thus
$$
L_{-1} = \pmatrix 0 & \lambda \\ 0 & 0 \endpmatrix\,, \qquad
L_0 = \pmatrix \half & 0 \\ 0 & -\half \endpmatrix\,, \qquad
L_{1} = \pmatrix 0 & 0 \\ -1 & 0 \endpmatrix\,. \eqno(22c)$$
In particular, it then follows that 
$$[L_m,L_n] = (m-n)L_{m+n}, \qquad m, n = 0, \pm 1\,.\eqno(23)$$
We also have that $L_n\Omega=0$, $n= 0, \pm 1$. With this
parametrisation, the operator corresponding to the M\"obius
transformation $\gamma$, defined in (19), is given as (see [13])
$$
U(\gamma) = \exp\left({b \over d} L_{-1}\right)\;
            \left(\sqrt{ ad -bc} \over d \right)^{L_0} \;
            \exp\left(-{c \over d} L_{1} \right)\,.
\eqno(24)$$

For $\psi\in V_h$, by (17), 
$U(\gamma)V(\psi,z)U(\gamma^{-1}) =V(\psi,\gamma(z))\gamma'(z)^h$,   
and so, by (21), 
$U(\gamma)\psi=\lim_{z\rightarrow 0}V(\psi,\gamma(z))\Omega
\gamma'(z)^h$. From this it follows that,   
$$L_0\psi=h\psi\,,\qquad L_1\psi=0\,, \qquad L_{-1}\psi =
V'(\psi,0)\Omega\,.\eqno(25)$$ 
Thus $L_0=\delta$ acting on $V$.

Henceforth we shall assume that our theory defined by $\A$ is M\"obius
invariant. 

Having chosen an identification of $\Pop$ with $\Cop\cup\{\infty\}$
and of $V$ with a subspace of $\V^\O$, we can now also define vertex
operators for $\bpsi=\prod_{j=1}^nV(\psi_j,z_j)\Omega\in\B_\C$ by 
$$V(\bpsi,z) = \prod_{j=1}^nV(\psi_j,z_j+z)\,.\eqno(26)$$
Then $V(\bphi,z)$ is a continuous operator
$\V^{\O_1}\rightarrow\V^{\O_2}$, provided that
$z_j+z\notin\O_2\subset\O_1$ but $z_j+z\in\O_1$, $1\leq j\leq n$.  

We can further extend the definition of $V(\bpsi,z)$ by linearity from
$\bpsi\in\B_\C$ to vectors $\Psi\in\sV_\C^\O$, the image of $\V_\C$ in
$\V_\C^\O$, to obtain a continuous linear operator 
$V(\Psi,z):\V^{\O_1} \rightarrow \V^{\O_2}$, where
$\C_z\cap\O_2=\empty$, $\O_2\subset\O_1$ and $\C_z\subset \O_1$ for  
$\C_z=\{\zeta+z :\zeta\in\C\}$. One might be tempted to try to extend
the definition of the vertex operator even further to states in
$\V^\O_\C\cong\V^\O$, but the corresponding operator will then only be
well-defined on a suitable dense subspace of $\V^{\O_1}$. 

For the vertex operator associated to $\Psi\in\sV_\C^\O$, we again
have  
$$e^{\lambda L_{-1}}V(\Psi,z) e^{-\lambda L_{-1}}
= V(\Psi,z+\lambda)\,, \qquad V(\Psi,0)\Omega = \Psi\,.\eqno(27)$$
Furthermore,
$$\eqalignno{
V(\Psi,z)V(\phi,\z) & = V(\phi,\z)V(\Psi,z)\,, & (28a) \cr
V(\Psi,z)\Omega & = e^{z L_{-1}}\Psi & (28b) } $$
for any $\phi\in V$, $\zeta\notin\C_z$. [In (28a), the lefthand and
righthand sides are to be interpreted as maps
$\V^{\O_1}\rightarrow\V^{\O_2}$, with  
$V(\phi,\z):\V^{\O_1}\rightarrow\V^{\O_L}$ and 
$V(\Psi,z):\V^{\O_L}\rightarrow\V^{\O_2}$ on the lefthand side and  
$V(\Psi,z): \V^{\O_1}\rightarrow\V^{\O_R}$ and 
$V(\phi,\z):\V^{\O_R}\rightarrow\V^{\O_2}$
on the righthand side, where $\O_2\subset\O_L\subset\O_1$,
$\O_2\subset\O_R\subset\O_1$, $\zeta\in\O_R\cap\O_L^c\cap\O_2^c$ and
$\C_z\subset\O_L\cap\O_R^c\cap\O_2^c$ (where $\O_2^c$ denotes the
complement of $\O_2$, {\it etc.}). Equation (28b) holds in $\V^\O$
with $\C_z\cap\O=\empty$.] 

Actually, these two conditions characterise the vertex operator
already uniquely: 

{\bf Theorem 3 [Uniqueness]:} For $\Psi\in \sV_\C^\O$, the
operator $V(\Psi,z)$ is uniquely characterised 
by the conditions (28a) and (28b). 

The proof is essentially that contained in ref.~[13]: If 
$W(z)V(\phi,\zeta)=V(\phi,\zeta)W(z)$ for $\phi\in V,\zeta\notin\C_z$,
and $W(z)\Omega = e^{zL_{-1}}\Psi$, it follows that, for 
$\Phi\in \sV_{\C'}^{\O'}$, $W(z)V(\Phi,\zeta)=V(\Phi,\zeta)W(z)$
provided that $\C_\zeta'\cap\C_z=\empty$ and so  
$$\eqalignno{
W(z)e^{\zeta L_{-1}}\Phi
=W(z)V(\Phi,\zeta)&\Omega=V(\Phi,\zeta)W(z)\Omega
=V(\Phi,\zeta)e^{z L_{-1}}\Psi\cr
&=V(\Phi,\zeta)V(\Psi,z)\Omega=V(\Psi,z)V(\Phi,\zeta)\Omega
=V(\Psi,z)e^{\zeta L_{-1}}\Phi}$$
for all $\Phi\in \sV_{\C'}^{\O'}$, which is dense in $\V^{\O'}$,
showing that $W(z)=V(\Psi,z)$.

{}From this uniqueness result and (17) we can deduce the commutators of
vertex operators $V(\psi,z)$, $\psi\in V_h$, with $L_{-1},L_0,L_1$:
$$\eqalignno{
[L_{-1}, V(\psi,z)] &= {d\over dz}V(\psi,z) &(29a)\cr
[L_0,V(\psi,z)]&= z{d\over dz}V(\psi,z) +h V(\psi,z) &(29b)\cr
[L_1,V(\psi,z)]&= z^2{d\over dz}V(\psi,z) +2hz V(\psi,z). &(29c)\cr
}$$

We recall from (25) that $L_1\psi=0$ and $L_0\psi=h\psi$ if 
$\psi\in V_h$; if $L_1\psi=0$, $\psi$ is said to be 
{\it quasi-primary}. 

The definition (26) immediately implies that, for states $\psi,\phi\in V$,
$V(\psi,z)V(\phi,\zeta)=V(V(\psi,z-\zeta)\phi,\zeta)$. This
statement generalises to the key duality result of Theorem 4, which can
be seen to follow from the uniqueness theorem:  

{\bf Theorem 4 [Duality]:} If $\Psi\in\sV_{\C}^\O$ and
$\Phi\in\sV_{\Cp}^\Op$, where $\C_z\cap\Cp_{\zeta}=\empty$, then 
$$V(\Psi,z)V(\Phi,\zeta) = V(V(\Psi,z-\zeta)\Phi,\zeta)\,.\eqno(30)$$

[In (30), the lefthand and righthand sides are to be interpreted as
maps $\V^{\O_1}\rightarrow\V^{\O_2}$, with
$V(\Phi,\z):\V^{\O_1}\rightarrow\V^{\O_L}$ and 
$V(\Psi,z): \V^{\O_L}\rightarrow\V^{\O_2}$ on the lefthand side and  
$V(\Psi,z-\zeta)\Phi\in \sV_{\C_{z-\zeta}\cup\Cp}$ where 
$\O_2\subset\O_L\subset\O_1$, $\C_z\subset \O_L\cap\O_2^c$ and
$\Cp_{\zeta}\subset\O_1\cap\O^c_L$.] 

The result follows from the uniqueness theorem on noting that 
$$V(\Phi,z)V(\Psi,\z)\Omega=V(\Phi,z)e^{\z L_{-1}}\Psi=
e^{\z L_{-1}}V(\Phi,z-\z)\Psi =V(V(\Phi,z-\z)\Psi),\z)\Omega.$$

\vskip24pt
{\bf 4. Modes, Fock Spaces and the Equivalence of Theories}
\vskip6pt

The concept of equivalence between two meromorphic field
theories in our definition could be formulated in terms of the whole
collection of spaces $\V^\O$, where $\O$ ranges over the open subsets
of $\rs$ with path-connected complement, but this would be very
unwieldy. In fact, each meromorphic field theory has a Fock
space at its heart and we can focus on this in order to define (and,
in practice, test for) the equivalence of theories. To approach this
we first need to introduce the concept of the modes of a vertex
operator. 

It is straightforward to see that we can construct contour integrals
of vectors in $\V^\O$, {\it e.g.} of the form
$$\int_{\C_1}dz_1\int_{\C_2}dz_2\ldots\int_{\C_r}dz_r
\mu(z_1,z_2,\ldots,z_r)\prod_{i=1}^n V(\psi_i,z_i)\Omega\,,
\eqno(31)$$
where $r\leq n$ and the weight function $\mu$ is analytic in some
neighbourhood of $C_1\times C_2\times\cdots\times C_r$ and the
distances $|z_i-z_j|$, $i\ne j$, are bounded away from $0$ on this
set. In this way we can define the modes 
$$V_n(\psi) = \oint_C z^{h+n-1}V(\psi,z)dz\,,\qquad \hbox{for } 
\psi\in V_h\,, \eqno(32)$$ 
as linear operators on $\V^\O_\C$, where $C$ encircles $\C$ and
$C\subset\O$ with $\infty\in\O$ and $0\in\C$, and we absorb a factor
of $1/2\pi i$ into the definition of the symbol $\oint$. The
meromorphicity of the amplitudes allows us to establish 
$$V(\psi,z) = \sum_{n=-\infty}^\infty V_n(\psi) z^{-n-h}\eqno(33)$$
with convergence with respect to the topology of $\V^{\O'}$ for
an appropriate $\O'$. 

The definition of $V_n(\psi)$ is independent of $C$ if it
is taken to be a simple contour encircling the origin once positively.
Further, if $\O_2\subset\O_1$, $\V^{\O_1}\subset\V^{\O_2}$ and if
$\infty\in\O_2, 0\notin\O_1$, the definition of $V_n(\psi)$ on
$\V^{\O_1}$, $\V^{\O_2}$, agrees on $\V^{\O_1}$, which is dense in
$\V^{\O_2}$, so that we may regard the definition as independent of
$\O$ also. $V_n(\psi)$ depends on our choice of $0$ and $\infty$ but
different choices can be related by M\"obius transformations.  

We define the Fock space $\H^\O\subset\V^\O$ to be the space spanned
by finite linear combinations of vectors of the form 
$$\Psi = V_{n_1}(\psi_1)V_{n_2}(\psi_2)\cdots
V_{n_N}(\psi_N)\Omega\,,\eqno(34)$$ 
where $\psi_j\in V$ and $n_j\in\Zop$, $1\leq j\leq N$. Then, by
construction, $\H^\O$ has a countable basis. It is easy to see that
$\H^\O$ is dense in $\V^\O$. Further it is clear that $\H^\O$ is
independent of $\O$, and, where there is no ambiguity, we shall denote
it simply by $\H$. It does however depend on the choice of $0$ and 
$\infty$, but different choices will be related by the action of the
M\"obius group again.  

It follows from (28b) that 
$$V(\psi,0)\Omega=\psi \eqno(35)$$
which implies that
$$V_n(\psi)\Omega = 0 \qquad\hbox{if } n>-h\eqno(36)$$
and 
$$V_{-h}(\psi)\Omega= \psi\,.\eqno(37)$$
Thus $V\subset\H$.

Since $\infty$ and $0$ play a special role, it is not surprising that
$L_0$, the generator of the subgroup of $\M$ preserving them, does as
well. From (29b) it follows that 
$$[L_0,V_n(\psi)]=-nV_n(\psi)\,,\eqno(38)$$
so that for $\Psi$ defined by (34),
$$L_0\Psi=h\Psi, \qquad\hbox{where } h = - \sum_{j=1}^N n_j\,.
\eqno(39)$$ 
Thus
$$\H=\bigoplus_{h\in\Zop}\H_h, \qquad \hbox{where } V_h\subset
\H_h\,,\eqno(40)$$ 
where $\H_h$ is the subspace spanned by vectors of the form (34) for
which $h=\sum_jn_j$. 

Thus $L_0$ has a spectral decomposition and the $\H_h$, $h\in\Zop$,
are the eigenspaces of $L_0$. They have countable dimensions but here
we shall only consider theories for which their dimensions are finite.
(This is not guaranteed by the finite-dimensionality of $V$; in fact,
in practice, it is not easy to determine whether these spaces are
finite-dimensional or not, although it is rather obvious in many
examples.) 

We can define vertex operators for the vectors (34) by
$$V(\Psi,z)=\oint_{\C_1}z_1^{h_1+n_1-1}V(\psi_1,z+z_1)dz_1\cdots
\oint_{\C_N}z_N^{h_N+n_N-1}V(\psi_N,z+z_N)dz_N\,,
\eqno(41)$$
where the $\C_j$ are contours about $0$ with $|z_i|>|z_j|$ if
$i<j$. We can then replace the densities (1) by the larger class $\A'$
of densities  
$$\langle V(\Psi_1,z_1)V(\Psi_2,z_2)\cdots V(\Psi_n,z_n)\rangle
\prod_{j=1}^n\left(dz_j\right)^{h_j}\,,
\eqno(42)$$
where $\Psi_j\in\H_{h_j}$, $1\leq j\leq n$. It is not difficult to see
that replacing $\A$ with $\A'$, {\it i.e.} replacing $V$ with $\H$,
does not change the definition of the spaces $\V^\O$. Theorem 3
(Uniqueness) and Theorem 4 (Duality) will still hold if we replace
$\sV^\O_\C$ with $\sH^\O_\C$, the space we would obtain if we started
with $\H$ rather than $V$, {\it etc.} These theorems enable the
M\"obius transformation properties of vertex operators to be
determined (see Appendix D). 

However, if we use the whole of $\H$ as a starting point, the M\"obius
properties of the densities $\A'$ can not be as simple as in  Section
3 because not all $\psi\in\H$ have the {\it quasi-primary}
property $L_1\psi=0$. But we can introduce the subspaces of
quasi-primary vectors within $\H$ and $\H_h$, 
$$\H^Q =\{\Psi\in\H : L_1\Psi=0\},\qquad \H_h^Q=\{\Psi\in\H_h :
L_1\Psi=0\},\qquad \H^Q=\bigoplus_h\H^Q_h\,.\eqno(43)$$
$V\subset\H^Q$ and $\H^Q$ is the maximal $V$ which will generate the
theory with the same spaces $\V^\O$ and with agreement of the
densities. [Under the cluster decomposition assumption of Section 6,
$\H$ is generated from $\H^Q$ by the action of the M\"obius group or,
more particularly, $L_{-1}$. See Appendix D.] 

We are now in a position to define the equivalence of two theories. A
theory specified by a space $V$ and amplitudes $\A=\{f\}$, leading to
a quasi-primary space $\H^Q$, is said to be equivalent to the theory
specified by a space $\hat V$ and amplitudes $\hat\A=\{\hat f\}$,
leading to a quasi-primary space $\hat\H^Q$, if there are graded
injections $\iota:V\rightarrow\hat\H^Q$ ({\it i.e.}
$\iota(V_h)\subset\hat\H_h^Q$) and $\hat\iota:\hat V\rightarrow\H^Q$ which
map amplitudes to amplitudes. 

Many calculations in conformal field theory are most easily performed
in terms of modes of vertex operators which capture in essence the
algebraic structure of the theory. In particular, the modes of the
vertex operators define what is usually called a W-algebra; this can
be seen as follows. 

The duality property of the vertex operators can be rewritten in terms
of modes as 
$$\eqalign{
V(\Phi,z)V(\Psi,\zeta) &= V(V(\Phi,z-\zeta)\Psi,\zeta)\cr
&=\sum_{n} V(V_n(\Phi)\Psi,\zeta) (z-\zeta)^{-n-h_\phi}\,,\cr}
\eqno(44)$$  
where $L_0\Psi=h_\Psi\Psi$ and $L_0\Phi=h_\Phi\Phi$, and 
$\Psi, \Phi\in\H$. We can then use the usual contour techniques of
conformal field theory to derive from this formula commutation
relations for the respective modes. Indeed, the commutator of two
modes $V_m(\Phi)$ and $V_n(\Psi)$ acting on 
$\B_\C$ 
is defined by
$$\eqalign{
[V_m(\Phi),\,V_n(\Psi)] & =
{\oint dz\oint d\zeta}_{\hskip-42pt
{\atop {\atop {\atop \scriptstyle |z|>|\zeta|}}}\hskip16pt}
z^{m+h_\Phi-1}
\zeta^{n+h_\Psi-1} V(\Phi,z) V(\Psi,\zeta) \cr & \qquad
-
{\oint dz\oint d\zeta}_{\hskip-42pt
{\atop {\atop {\atop \scriptstyle |\zeta|>|z|}}}\hskip16pt}
z^{m+h_\Phi-1} \zeta^{n+h_\Psi-1} V(\Phi,z) V(\Psi,\zeta)\,,} 
\eqno(45)$$
where the contours on the right-hand side encircle $\C$
anti-clockwise. We can then deform the two contours so as to rewrite
(45) as  
$$[V_m(\Phi),\,V_n(\Psi)]=\oint_0 \zeta^{n+h_\Psi-1}d\zeta\, 
\oint_\zeta z^{m+h_\Phi-1} dz\,
\sum_{l} V(V_l(\Phi)\Psi,\zeta) (z-\zeta)^{-l-h_\Phi} \,,\eqno(46)$$ 
where the $z$ contour is a small positive circle about $\zeta$ and the
$\zeta$ contour is a positive circle about $\C$. Only terms with 
$l\geq 1-h_\Phi$ contribute, and the integral becomes
$$
[V_m(\Phi),V_n(\Psi)] = \sum_{N=-h_\Phi+1}^{\infty}
\pmatrix m+h_\Phi-1\\ m-N\endpmatrix V_{m+n}(V_N(\Phi)\Psi)\,.
\eqno(47)$$
In particular, if $m\geq-h_\Phi+1$, $n\geq-h_\Psi+1$, 
then $m-N\geq0$ in the sum, and $m+n\geq N+n\geq N-h_\Psi+1$. This
implies that the modes $\left\{V_m(\Psi) :m\geq-h_\Psi+1\right\}$ 
close as a Lie algebra; the same also holds for 
$\left\{V_m(\Psi) : 0\geq m \geq -h_\Psi + 1\right\}$.

As we shall discuss below in Section 6, in conformal field theory it is
usually assumed that the amplitudes satisfy another property which
guarantees that the spectrum of $L_0$ is bounded below by $0$. If this
is the case then the sum in (47) is also bounded above by $h_\Psi$.

\vskip24pt
{\bf 5. Some Examples}
\vskip6pt

Before proceeding further, we shall give a number of examples of theories
that satisfy the axioms that we have specified so far.
\vskip8pt

{\bf (a) The $U(1)$ theory}
\vskip6pt

The simplest example is the case where $V$ is a
one-dimensional vector space, spanned by a vector $J$ of weight
$1$, in which case we write $J(z)\equiv V(J,z)$. The amplitude of an
odd number of $J$-fields is defined to vanish, and in the case of an even
number it is given by
$$\eqalignno{
\langle J(z_1) \cdots J(z_{2n}) \rangle &= {k^n \over 2^n n!} 
\sum_{\pi\in S_{2n}} \,
\prod_{j=1}^{n} {1 \over (z_{\pi(j)} - z_{\pi(j+n)})^2} \,,&(48a)\cr
&= k^n
\sum_{\pi\in S_{2n}'} \,
\prod_{j=1}^{n} {1 \over (z_{\pi(j)} - z_{\pi(j+n)})^2} \,,&(48b)\cr
}$$
where $k$ is an arbitrary (real) constant and, in (48a), $S_{2n}$ is
the permutation group on $2n$ object, whilst, in (48b), the sum is
restricted to the subset $S_{2n}'$ of permutations $\pi\in S_{2n}$
such that $\pi(i)<\pi(i+n)$ and $\pi(i)<\pi(j)$ if $1\leq i<j\leq n$. 
(This defines the amplitudes on a basis of $V$ and we extend the
definition by multilinearity.) It is clear that the amplitudes are
meromorphic in $z_j$, and that they satisfy the locality condition. It
is also easy to check that they are M\"obius covariant, with the
weight of $J$ being $1$.  

{}From the amplitudes we can directly read off the operator product
expansion of the field $J$ with itself as 
$$
J(z) J(w) \sim {k \over (z-w)^2} + O(1) \,. \eqno(49)
$$
Comparing this with (44), and using (47) we then obtain
$$
[J_n,J_m] = n k \delta_{n,-m} \,.\eqno(50)
$$
This defines (a representation of) the affine algebra $\hat{u}(1)$.
\vskip8pt

{\bf (b) Affine Lie algebra theory}
\vskip6pt

Following Frenkel and Zhu [20], we can generalise this example to the
case of an arbitrary finite-dimensional Lie algebra $g$. Suppose that
the matrices $t^a$, $1\leq a\leq \dim g$, provide a finite-dimensional
representation of $g$ so that $[t^a,t^b]={f^{ab}}_ct^c$, where
${f^{ab}}_c$ are the structure constants of $g$. In this case, the
space $V$ is of dimension $\dim g$ and has a basis consisting of
weight one states $J^a$, $1\leq a\leq \dim g$. Again, we write
$J^a(z)=V(J^a,z)$.  

If $K$ is any matrix which commutes with all the $t^a$, define 
$$\kappa^{a_1a_2\ldots a_m}
=\hbox{tr}(Kt^{a_1}t^{a_2}\cdots t^{a_m})\,.\eqno(51)$$ 
The $\kappa^{a_1a_2\ldots a_m}$ have the properties that
$$\kappa^{a_1a_2a_3\ldots a_{m-1}a_m}
=\kappa^{a_2a_3\ldots a_{m-1}a_ma_1} \eqno(52)$$ 
and
$$\kappa^{a_1a_2a_3\ldots a_{m-1}a_m}
-\kappa^{a_2a_1a_3\ldots a_{m-1}a_m}=
{f^{a_1a_2}}_b \kappa^{ba_3\ldots a_{m-1}a_m}\,.\eqno(53)$$
With a cycle $\sigma=(i_1,i_2,\ldots,i_m)\equiv  (i_2,\ldots,i_m,i_1)$
we associate the function
$$f_\sigma^{a_{i_1}a_{i_2}\ldots
a_{i_m}}(z_{i_1},z_{i_2},\ldots,z_{i_m}) 
= {\kappa^{a_{i_1}a_{i_2}\ldots a_{i_m}}
\over (z_{i_1}-z_{i_2})(z_{i_2}-z_{i_3})\cdots (z_{i_{m-1}}-z_{i_m})
(z_{i_m}-z_{i_1})}\,.\eqno(54)
$$
If the permutation $\rho\in S_n$ has no fixed points, it can be
written as the product of cycles of length at least 2,
$\rho=\sigma_1\sigma_2\ldots\sigma_M$. We associate to $\rho$ the
product $f_\rho$ of functions 
$f_{\sigma_1}f_{\sigma_2}\ldots f_{\sigma_M}$ and 
define $\langle J^{a_1}(z_1)J^{a_2}(z_2)\ldots J^{a_n}(z_n)\rangle$ to
be the sum of such functions $f_\rho$ over permutations $\rho\in S_n$
with no fixed point. Graphically, we can construct these amplitudes by
summing over all graphs with $n$ vertices where the vertices carry
labels $a_j$, $1\leq j\leq n$, and each vertex is connected by two
directed lines (propagators) to other vertices, one of the lines at
each vertex pointing towards it and one away. Thus, in a given graph,
the vertices are divided into directed loops or cycles, each loop
containing at least two vertices. To each loop, we associate a
function as in (55) and to each graph we associate the product of
functions associated to the loops of which it is composed. 

Again, this defines the amplitudes on a basis of $V$ and we extend the
definition by multilinearity. The amplitudes are evidently local and
meromorphic, and one can verify that they satisfy the M\"obius
covariance property with the weight of $J^a$ being $1$.

The amplitudes determine the operator product expansion to
be of the form
$$
J^a(z) J^b(w) \sim {\kappa^{ab} \over (z-w)^2} 
+ {{f^{ab}}_c J^c(w)\over (z-w)}  + O(1) \,, \eqno(55)
$$
and the algebra therefore becomes
$$
[J^a_m,J^b_n] = {f^{ab}}_c J^c_{m+n} + m \kappa^{ab} \delta_{m,-n}
\,.\eqno(56)
$$
This is (a representation of) the affine algebra $\hat{g}$. In the
particular case where $g$ is simple,
$\kappa^{ab}=\hbox{tr}(Kt^at^b)=k\delta^{ab}$, for some $k$, if  
we choose a suitable basis.
\vskip8pt

{\bf (c) The Virasoro Theory}
\vskip6pt

Again following Frenkel and Zhu [20], we can construct the Virasoro
theory in a similar way. In this case, the space $V$ is
one-dimensional, spanned by a vector $L$ of weight $2$ and we write
$L(z)=V(L,z)$.  We can again construct the amplitudes graphically by
summing over all graphs with $n$ vertices, where the vertices are
labelled by the integers $1\leq j\leq n$, and each vertex is connected
by two lines (propagators) to other vertices. In a given graph, the
vertices are now divided into loops, each loop containing of at least
two vertices. To each loop $\ell=(i_1,i_2,\ldots,i_m)$, we associate a
function
$$f_\ell(z_{i_1},z_{i_2},\ldots,z_{i_m}) = {c/2
\over (z_{i_1}-z_{i_2})^2(z_{i_2}-z_{i_3})^2\cdots 
(z_{i_{m-1}}-z_{i_m})^2 (z_{i_m}-z_{i_1})^2},\eqno(57)
$$
where $c$ is a real number, and, to a graph, the product of the functions
associated to its loops. [Since it corresponds a factor of the form
$(z_i-z_j)^{-2}$ rather than $(z_i-z_j)^{-1}$, each line or propagator
might appropriately be represented by a double line.] The amplitudes
$\langle L(z_1)L(z_2)\ldots L(z_n)\rangle$ are then obtained by summing the
functions associated with the various graphs with $n$ vertices. [Note
graphs related by reversing the direction of any loop contribute equally to
this sum.]

These amplitudes determine the operator product expansion to be 

$$L(z)L(\zeta)\sim {c/2\over (z-\zeta)^4} + {2L(\zeta)\over (z-\zeta)^2}
+{L'(\zeta)\over z-\zeta} + O(1)\eqno(58)
$$
which leads to the Virasoro algebra
$$[L_m,L_n]=(m-n)L_{m+n}+{c\over12}m(m^2-1)\delta_{m,-n}\,.
\eqno(59)$$ 
\vskip8pt

{\bf (d) Lattice Theories}
\vskip6pt

Suppose that $\Lambda$ is an even $n$-dimensional Euclidean lattice, so
that, if $k\in\Lambda$, $k^2$ is an even integer. We introduce a basis 
$e_1,e_2,\ldots, e_n$ for $\Lambda$, so that any element $k$ of
$\Lambda$ is an integral linear combination of these basis
elements. We can introduce an algebra consisting of 
matrices $\gamma_j$, $1\leq j\leq n$, such that $\gamma_j^2=1$ and
$\gamma_i\gamma_j = (-1)^{e_i\cdot e_j}\gamma_j\gamma_i$. If we define 
$\gamma_k=\gamma_1^{m_1}\gamma_2^{m_2}\ldots\gamma_n^{m_n}$ for
$k=m_1e_1+m_2e_2+\ldots+m_ne_n$, we can define quantities
$\epsilon(k_1,k_2,\ldots, k_N)$, taking the values $\pm 1$, by 
$$\gamma_{k_1}\gamma_{k_2}\ldots\gamma_{k_N}=
\epsilon(k_1,k_2,\ldots, k_N)\gamma_{k_1+k_2+\ldots+k_N}\,.\eqno(60)
$$
We define the theory associated to the lattice $\Lambda$ by taking $V$ to
have a basis $\{\psi_k:k\in\Lambda\}$, where the weight of $\psi_k$ is
$\half k^2$, and, writing $V(\psi_k,z)=V(k,z)$, the amplitudes to be 
$$\langle V(k_1,z_1)V(k_2,z_2)\cdots V(k_N,z_n)\rangle
= \epsilon(k_1,k_2,\ldots, k_N)\prod_{1\leq i<j\leq
N}(z_i-z_j)^{k_i\cdot k_j}\eqno(61)$$
if $k_1+k_2+\ldots+ k_N=0$ and zero otherwise. The
$\epsilon(k_1,k_2,\ldots, k_N)$ obey the conditions 
$$\eqalign{
\epsilon(k_1,k_2,\ldots,k_{j-1},k_j,&k_{j+1},k_{j+2},\ldots, k_N)\cr 
&= (-1)^{k_j\cdot k_{j+1}}
\epsilon(k_1,k_2,\ldots,k_{j-1},k_{j+1},k_j,k_{j+2},\ldots,k_N),  
}$$
which guarantees locality, and 
$$\eqalign{
\epsilon(k_1,k_2,\ldots&, k_j)\epsilon(k_{j+1},\ldots,
k_N)\cr
&=\epsilon(k_1+k_2+\ldots+ k_j,k_{j+1}+\ldots+
k_N)\epsilon(k_1,k_2,\ldots,k_j,k_{j+1},\ldots, k_N),\cr}$$
which implies the cluster decomposition property of Section 6.

\vskip8pt

{\bf (e) A non-conformal example}
\vskip6pt

The above examples actually define meromorphic {\it conformal} field
theories, but since we have not yet defined what we mean by a theory
to be conformal, it is instructive to consider also an example that
satisfies the above axioms but is not conformal. The simplest such
case is a slight modification of the $U(1)$ example described in 5(a): 
again we take $V$ to be a one-dimensional vector space, spanned by a
vector $K$, but now the grade of $K$ is taken to be $2$. Writing 
$K(z)\equiv V(K,z)$, the amplitudes of an odd number of $K$-fields
vanishes, and in the case of an even number we have
$$
\langle K(z_1) \cdots K(z_{2n}) \rangle = {k^n \over 2^n n!} 
\sum_{\pi\in S_{2n}}\,
\prod_{j=1}^{n} {1 \over (z_{\pi(j)} - z_{\pi(j+n)})^4} \,.\eqno(62)
$$
It is not difficult to check that these amplitudes satisfy all the
axioms we have considered so far. In this case the operator product
expansion is of the form 
$$
K(z) K(w) \sim {k \over (z-w)^4} + O(1) \,,\eqno(63)
$$
and the algebra of modes is given by
$$
[K_n,K_m] = {k \over 6} n (n^2-1) \delta_{n,-m} \,. \eqno(64)
$$

\vskip24pt
{\bf 6. Cluster Decomposition}
\vskip6pt

So far the axioms we have formulated do not impose any restrictions on
the relative normalisation of amplitudes involving for example a
different number of vectors in $V$, and the class of theories we are
considering is therefore rather flexible. This is mirrored by the fact
that it does not yet follow from our considerations that the spectrum
of the operator $L_0$ is bounded from below, and since $L_0$ is in
essence the energy of the corresponding physical theory, we may want
to impose this constraint. In fact, we would like to impose the
slightly stronger condition that the spectrum of $L_0$ is bounded by
$0$, and that there is precisely one state with eigenvalue equal to
zero. This will follow (as we shall show momentarily) from the 
{\it cluster decomposition property}, which states that if we separate
the variables of an amplitude into two sets and scale one set towards
a fixed point ({\it e.g.} $0$ or $\infty$) the behaviour of the
amplitude is dominated by the product of two amplitudes, corresponding
to the two sets of variables, multiplied by an appropriate power of
the separation, specifically 
$$\left\langle\prod_i V(\phi_i,\zeta_i)\prod_j V(\psi_j,\lambda
z_j)\right\rangle
\sim \left\langle\prod_i V(\phi_i,\zeta_i)\right\rangle
\left\langle\prod_j V(\psi_j,z_j)\right\rangle \lambda^{-\Sigma h_j}
\qquad \hbox{as } \lambda\rightarrow 0\,,\eqno(65)$$
where $\phi_i\in V_{h_i'}, \psi_j\in V_{h_j}$. 
It follows from  M\"obius invariance, that
this is equivalent to
$$\left\langle\prod_i V(\phi_i,\lambda\zeta_i)\prod_j V(\psi_j,
z_j)\right\rangle
\sim \left\langle\prod_i V(\phi_i,\zeta_i)\right\rangle
\left\langle\prod_j V(\psi_j,z_j)\right\rangle \lambda^{-\Sigma h_i'}
\qquad \hbox{as } \lambda\rightarrow \infty\,.\eqno(66)$$
The cluster decomposition property extends also to vectors
$\Phi_i,\Psi_j\in \H$. It is not difficult to check that the examples
of the previous section satisfy this condition.

We can use the cluster decomposition property to show that the
spectrum of $L_0$ is non-negative and that the vacuum is, in a sense,
unique. To this end let us introduce the projection operators defined
by  
$$P_N = \oint_0 u^{L_0-N-1}du, \qquad\hbox{for } N\in\Zop\,.\eqno(67)$$
In particular, we have
$$P_N \prod_j V(\psi_j,z_j)\Omega = \oint u^{h - N -1}
V(\psi_j,uz_j)\Omega du \,,\eqno(68)$$
where $h=\sum_j h_j$. It then follows that the $P_N$ are projection
operators
$$P_NP_M=0,\hbox{ if } N\ne M, \qquad P_N^2=P_N, \qquad 
\sum_N P_N =1\eqno(69)$$ 
onto the eigenspaces of $L_0$,
$$L_0P_N=NP_N\,.\eqno(70)$$ 
For $N\leq 0$, we then have
$$\eqalignno{
\left\langle\prod_i V(\phi_i,\zeta_i)P_N\prod_j V(\psi_j,z_j)
\right\rangle
&=\oint_0 u^{\Sigma h_j- N-1}
\left\langle\prod_i V(\phi_i,\zeta_i)\prod_j V(\psi_j,u z_j)
\right\rangle du\cr
&\sim \left\langle\prod_i V(\phi_i,\zeta_i)\right\rangle
\left\langle\prod_j V(\psi_j,z_j)\right\rangle \oint_{|u|=\rho}
u^{-N-1}du ,}$$
which, by taking $\rho\rightarrow 0$, is seen to vanish for $N<0$ and,
for $N=0$, to give 
$$P_0\prod_j V(\psi_j,z_j)\Omega 
=\Omega\left\langle\prod_j V(\psi_j,z_j)\right\rangle\,,\eqno(71)$$ 
and so $P_0\Psi = \Omega\,\langle\Psi\rangle$. Thus the cluster 
decomposition property implies that $P_N=0$ for $N<0$, {\it i.e.}
the spectrum of $L_0$ is non-negative, and that $\H_0$ is spanned by
the vacuum $\Omega$, which is thus the unique state with $L_0=0$. 

As we have mentioned before the absence of negative eigenvalues of
$L_0$ gives an upper bound on the order of the pole in the operator
product expansion of two vertex operators, and thus to an upper bound
in the sum in (47): if $\Phi, \Psi\in\H$ are of grade
$L_0\Phi=h_\Phi\Phi, L_0\Psi=h_\Psi\Psi$, 
we have that $V_n(\Phi)\Psi =0$ for $n>h_\Psi$ because otherwise
$V_n(\Phi)\Psi$ would have a negative eigenvalue, $h_\Psi-n$, with
respect to $L_0$. In particular, this shows that the leading
singularity in $V(\Phi,z)V(\Psi,\zeta)$ is at most
$(z-\zeta)^{-h_\Psi-h_\Phi}$.

The cluster property also implies that the space of states of the
meromorphic field theory does not have any proper invariant subspaces
in a suitable sense. To make this statement precise we must first give
a meaning to a subspace of the space of states of a conformal field
theory. The space of states of the theory is really the collection of
topological spaces $\V^\O$, where $\O$ is an open subset of $\rs$
whose complement is path-connected. Recall that $\V^\O\subset
\V^{\O'}$ if $\O\supset \O'$. By a subspace of the conformal field
theory we shall mean subspaces $\U^\O\subset\V^\O$ specified for each
open subset $\O\subset\rs$ with path-connected complement, such that
$\U^\O=\U^{\O'}\cap\V^\O$ if $\O\supset \O'$.

{\bf Proposition 5:} Suppose $\{\U^\O\}$ is an invariant closed
subspace of $\{\V^\O\}$, {\it i.e.} $\U^\O$ is closed;
$\U^\O=\U^{\O'}\cap\V^\O$ if $\O\supset \O'$; and
$V(\psi,z)\U^\O\subset\U^{\O'}$ for all $\psi\in V$ where 
$z\in\O$, $z\notin\O'\subset\O$. Then $\{\U^\O\}$ is not a proper
subspace, {\it i.e.} either $\U^\O=\V^\O$ for all $\O$, or
$\U^\O=\{0\}$.   

{\bf Proof:} Suppose that $\phi\in\U^\O$, $\psi_j\in V$, $z_j\in\O$,
$z_j\notin\O'\subset\O$ and consider 
$$
\prod_{j=1}^n V(\psi_j,z_j)\phi \in\U^{\O'}\,.\eqno(72)$$
Now, taking a suitable integral of the left hand side,
$$P_0 \prod_{j=1}^nV(\psi_j,z_j)\phi=\lambda\Omega
=\left\langle
\prod_{j=1}^nV(\psi_j,z_j)\phi\right\rangle\Omega\,.\eqno(73)$$
Thus either all the amplitudes involving $\phi$ vanish 
for all $\phi\in\U$, in which case $\U=\{0\}$, or  
$\Omega\in\U^{\O'}$ for some $\O'$, in which case it is easy to see that 
$\Omega\in\U^\O$ for all $\O$ and it follows that $\U^\O=\V^\O$ for all
$\O$.

The cluster property also implies that the image of $\B_\C$ in
$\V_\C^\O$ is faithful. To show that the images of the elements
$\bpsi, \bpsi'\in\B_\C$ are distinct we note that otherwise
$\eta_\bphi(\bpsi)=\eta_\bphi(\bpsi')$ for all $\bphi\in\V_\O$ with
$\bphi$ as in (4). By taking $m$ in (4) to be sufficiently large,
dividing the $\zeta_i$, $1\leq i\leq m$, into $n$ groups which we
allow to approach the $z_j$, $1\leq j\leq n$, successively.  The
cluster property then shows that these must be the same points as the
$z'_j$, $1\leq j\leq n'$ in $\bpsi'$ and that 
$\psi_i=\mu_j\psi'_j$ for some $\mu_j\in\Cop$ with
$\prod_{j=1}^n\mu_j=1$, establishing that $\bpsi=\bpsi'$ as elements
of $\V^\O_\C$.

\vskip24pt
{\bf 7. Conformal Symmetry}
\vskip6pt

So far our axioms do not require that our amplitudes correspond to a
conformal field theory, only that the theory have a M\"obius
invariance, and indeed, as we shall see, the example in 5(c) is not
conformally invariant.  Further, what we shall discuss in the sections
which follow the present one will not depend on a conformal structure,
except where we explicitly mention it; in this sense, the present
section is somewhat of an interlude.  On the other hand, the conformal
symmetry is crucial for more sophisticated considerations, in
particular the theory on higher genus Riemann surfaces, and therefore
forms a very important part of the general framework.

Let us first describe a construction be means of which a potentially
new theory can be associated to a given theory, and explain then in
terms of this construction what it means for a theory to be
conformal. 

Suppose we are given a theory that is specified by a space $V$ and
amplitudes $\A=\{f\}$. Let us denote by $\hat{V}$ the vector space
that is obtained from $V$ by appending a vector $L$ of grade two, and
let us write $V(L,z)=L(z)$. The amplitudes involving only fields in
$V$ are given as before, and the amplitude
$$
\langle \prod_{j=1}^{m} L(w_j) \prod_{i=1}^{n} V(\psi_i,z_i)
\rangle \,,\eqno(74)
$$
where $\psi_i\in V_{h_i}$ is defined as follows: we associate to each
of the $n+m$ fields a point, and then consider the (ordered) graphs
consisting of loops where each loop contains at most one of the points
associated to the $\psi_i$, and each point associated to an $L$ is a
vertex of precisely one loop. (The points associated to $\psi_i$ may be
vertices of an arbitrary number of loops.) To each loop whose vertices
only consist of points corresponding to $L$ we associate the same
function as before in Section 5(c), and to the loop
$(z_i,w_{\pi(1)}, \ldots, w_{\pi(l)})$ we associate the
expression 
$$\eqalignno{
\prod_{j=1}^{l-1} {1 \over (w_{\pi(j)} - w_{\pi(j+1)})^2}
&\left( {h_i \over (w_{\pi(1)} - z_i)  (w_{\pi(l)} - z_i)} \right.\cr
&\hskip20truemm+ \left.{1 \over 2} \left[ {1 \over (w_{\pi(1)} - z_i)} {d
\over dz_i} 
 + {1\over (w_{\pi(l)} - z_i)} {d \over dz_i} \right] \right)
\,. &(75)
}$$
We then associate to each graph the product of the expressions
associated to the different loops acting on the amplitude which is
obtained from (74) upon removing $L(w_1)\cdots L(w_m)$, and
the total amplitude is the sum of the functions associated to all such
(ordered) graphs. (The product of the expressions of the form (75) is
taken to be `normal ordered' in the sense that all derivatives with
respect to $z_i$ only act on the amplitude that is obtained from (74)
upon excising the $L$s; in this way, the product is independent of the
order in which the expressions of the from (75) are applied.) 

We extend this definition by multilinearity to amplitudes defined for
arbitrary states in $\hat{V}$. It follows immediately that the
resulting amplitudes are local and meromorphic; in Appendix~E we shall
give a more explicit formula for the extended amplitudes, and use it
to prove that the amplitudes also satisfy the M\"obius covariance and
the cluster property. In terms of conventional conformal field theory,
the construction treats all quasiprimary states in $V$ as primary
with respect to the Virasoro algebra of the extended theory; this is
apparent from the formula given in Appendix E.

We can generalise this definition further by considering in addition
graphs which contain `double loops' of the form $(z_i,w_j)$ for those 
points $z_i$ which correspond to states in $V$ of grade two, where in
this case neither $z_i$ nor $w_j$ can be a vertex of any other loop. 
We associate the function
$$
{c_{\psi}/2 \over (z_i-w_j)^4} \eqno(76)
$$
to each such loop (where $c_\psi$ is an arbitrary linear functional on
the states of weight two in $V$), and the product of the different
expressions corresponding to the different loops in the graph act in
this case on the amplitude (74), where in addition to all $L$-fields
also the fields corresponding to $V(\psi_i,z_i)$ (for each $i$ which
appears in a double loop) have been removed. It is easy to see that
this generalisation also satisfies all axioms.

This construction typically modifies the structure of the meromorphic
field theory in the sense that it changes the operator product
expansion (and thus the commutators of the corresponding modes) of
vectors in $V$; this is for example the case for the `non-conformal'
model described in Section 5(e). If we introduce the field $L$ 
as described above, we find the commutation relations 
$$ [L_m, K_n ] = (m-n)\, K_{m+n} 
+ {c_K \over 12}\, m \, (m^2 -1) \, \delta_{m,-n}\,.
\eqno(77) $$
However, this is incompatible with the original commutator in (64):
the Jacobi identity requires that 
$$
\eqalignno{
0 & = 
\left[L_m,[K_n,K_l]\,\right] + \left[K_n,[K_l,L_m]\,\right] + 
\left[K_l,[L_m,K_n]\,\right] \cr
& = 
(l-m) \; [K_n,K_{l+m}] + (m-n) \; [K_l,K_{m+n}] \cr
& = 
\frac{k}{6} \, \delta_{l+m+n,0} \left[ - (l-m)\;(l+m)\; 
\left((l+m)^2 - 1\right) 
+ (2 m + l) \;l \; (l^2 - 1) \right] \cr
& = 
\frac{k}{6} \, \delta_{l+m+n,0} \; m \; (m^2 - 1) \; (2l + m)}
$$
and this is not satisfied unless $k=0$ (in which case the original
theory is trivial). In fact, the introduction of $L$ modifies (64) as 
$$
\eqalignno{
[ K_m, K_n ] & =  {k \over 6} m (m^2 -1) \delta_{m,-n} + {k \over a}
(m-n) Z_{m+n}  \cr
\ [ L_m, K_n ] & =  {c_K \over 12} m (m^2 -1) \delta_{m,-n} + (m-n)
K_{m+n}  \cr
\ [ Z_m, K_n ] & = 0  \cr
\ [ L_m, Z_n ] & = {a \over 12} m (m^2 -1) \delta_{m,-n} + (m-n)
Z_{m+n}  \cr
\ [ L_m, L_n ] & = {c \over 12} m (m^2 -1) \delta_{m,-n}
+ (m-n) L_{m+n}  \cr
\ [ Z_m, Z_n ] & = 0 \,, }
$$
where $a$ is non-zero and can be set to equal $k$ by rescaling $Z$,
and the $Z_n$ are the modes of a field of grade two. This set of
commutators then satisfies the Jacobi identities. It also follows from
the fact that the commutators of $Z$ with $K$ and $Z$ vanish, that
amplitudes that involve only $K$-fields and at least one $Z$-field
vanish; in this way we recover the original amplitudes and
commutators.  

The construction actually depends on the choice of $V$ (as well as the
values of $c_\psi$ and $c$), and therefore does not only depend on the
equivalence class of meromorphic field theories. However, we can ask
whether a given equivalence class of meromorphic field theories
contains a representative $(V,\A)$ (\ie\ a choice of $V$ that gives an
equivalent description of the theory) for which $(\hat{V},\hat\A)$ is
equivalent to $(V,\A)$; if this is the case, we call the meromorphic
field theory {\it conformal}. It follows directly from the definition
of equivalence that a meromorphic field theory is conformal if and
only if there exists a representative $(V,\A)$ and a vector
$L^0\in V$ (of grade two) so that
$$
\langle \left( L(w) - L^0(w) \right) \prod_{j=1}^{n} V(\psi_j,z_j)
\rangle = 0 \eqno(78)
$$
for all $\psi_j\in V$, where $L$ is defined as above. In this case,
the linear functional $c_\psi$ is defined by 
$$ 
c_{\psi} = 2 (w-z)^4 \langle L^0(w) V(\psi,z) \rangle \,.
$$
In the case of the non-conformal example of Section 5(e), it is clear
that (78) cannot be satisfied as the Fock space only contains one
vector of grade two, $L^0= \alpha K_{-2} \Omega$, and $L^0$ does not
satisfy (78) for any value of $\alpha$. On the other hand, for the
example of Section 5(a), we can choose 
$$ 
L^0 = { 1\over 2k} \, J_{-1}J_{-1}\Omega\,,\eqno(79)
$$
and this then satisfies (78). Similarly, in the case of the example of
Section 5(b), we can choose
$$
L^0 = {1 \over 2(k+Q)} \sum_a J^a_{-1} J^a_{-1} \Omega \,,
$$
where $Q$ is the dual Coxeter number of $g$ (\ie\ the value of the
quadratic Casimir in the adjoint representation), and again (78) is
satisfied for this choice of $L^0$ (and the above choice of $V$). This
construction is known as the `Sugawara construction'.

For completeness it should be mentioned that the modes of the field
$L$ (that is contained in the theory in the conformal case) satisfy
the Virasoro algebra
$$
[L_m,L_n] = (m-n) L_{m+n} + {c \over 12} m (m^2 -1) \delta_{m,-n} \,,
$$
where $c$ is the number that appears in the above definition of
$L$. Furthermore, the modes $L_m$ with $m=0,\pm 1$ agree with the
M\"obius generators of the theory.

\vskip24pt
\leftline{\bf 8. Representations}
\vskip6pt

In order to introduce the concept of a representation of a meromorphic
conformal field theory or conformal algebra, we consider a collection
of densities more general than those used in Section 2 to define the
meromorphic conformal field theory itself. The densities we now
consider are typically defined on a cover of the Riemann sphere,
$\rs$, rather than $\rs$ itself. We consider densities which are
functions of variables $u_i$, $1\leq i\leq N$, and $z_j$, 
$1\leq j\leq n$, which are analytic if no two of these $N+n$ variables
are equal, may have poles at $z_i=z_j$, $i\ne j$, or $z_i=u_j$, and
may be branched about $u_i=u_j$, $i\ne j$. To define a representation, we
need the case where
$N=2$, in which the densities are meromorphic in all but two of the
variables.  

Starting again with $V=\oplus_h V_h$, together with two finite-dimensional
spaces $W_\alpha$ and $W_\beta$ (which may be one-dimensional), we suppose
that, for each  integer $n\geq 0$, and
$z_i\in\rs$ and $u_1,u_2$ on some branched cover of $\rs$, and for any
collection of vectors $\psi_i\in V_{h_i}$ and 
$\chi_1\in W_\alpha,\chi_2\in W_\beta$, we have a density
$$\eqalignno{
g(\psi_1&,\ldots,\psi_n;z_1, \ldots, z_n; \chi_1,\chi_2; u_1,u_2)\cr 
\equiv \,& \langle V(\psi_1,z_1)V(\psi_2,z_2)\cdots V(\psi_n,z_n)
W_\alpha(\chi_1,u_1) W_\beta(\chi_2,u_2)\rangle 
\prod_{j=1}^n\left(dz_j\right)^{h_j} 
(d u_1)^{r_1} (d u_2)^{r_2}\,,\cr
& &(80)} $$
where $r_1, r_2$ are real numbers, which we call the {\it conformal 
weights} of $\chi_1$ and $\chi_2$, respectively. The amplitudes
$$\langle V(\psi_1,z_1)V(\psi_2,z_2)\cdots V(\psi_n,z_n) 
W_\alpha(\chi_1,u_1) W_\beta(\chi_2,u_2)\rangle\eqno(81)$$
are taken to be multilinear in the $\psi_j$ and $\chi_1,\chi_2$,
and invariant under the exchange of $(\psi_i,z_i)$ with
$(\psi_j,z_j)$, and meromorphic as a function of the $z_j$, analytic
except for possible poles at $z_i=z_j$, $i\ne j$, and $z_i=u_1$ or
$z_i=u_2$. As functions of $u_1, u_2$, the amplitudes are analytic
except for the possible poles at $u_1=z_i$ or $u_2=z_i$ and a possible
branch cut at $u_1=u_2$. We denote a collection of such densities by
$\R=\{g\}$. 

Just as before, given an open set $\C\subset\rs$ we introduced spaces
$\B_\C$, whose elements are of the form (3), so we can now introduce 
sets, $\B_{\C\alpha\beta}$, labelled by  finite collections of
$\psi_i\in V_{h_i}$, $z_i\in\C\subset \rs$, $i = 1,\ldots, n$,
$n\in\Nop$ and $z_i\ne z_j$ if $i\ne j$, together with 
$\chi_1\in W_\alpha, \chi_2\in W_\beta$ and $u_1, u_2\in\C$, 
$u_1\ne u_2$ and $z_i\ne u_j$, denoted by 
$$\eqalignno{
\bchi &= V(\psi_1,z_1)V(\psi_2,z_2)\cdots V(\psi_n,z_n)
W_\alpha(\chi_1,u_1) W_\beta(\chi_2,u_2) \Omega \cr
&\equiv \prod_{i=1}^n V(\psi_i,z_i)W_\alpha(\chi_1,u_1)
W_\beta(\chi_2,u_2) \Omega\,.&(82)\cr}$$ 
We again immediately identify different $\bchi\in\B_{\C\alpha\beta}$
with the other elements of $\B_{\C\alpha\beta}$ obtained by replacing
each $\psi_j$ in (82) by $\mu_j\psi_j$, $1\leq j\leq n$, $\chi_i$ by 
$\lambda_i\chi_i$, $i=1,2$, where $\lambda_1, \lambda_2, \mu_j\in\Cop$
and $\lambda_1\lambda_2\prod_{j=1}^n\mu_j=1$.

Proceeding as before, we introduce the vector space
$\V_{\C\alpha\beta}$ with basis $\B_{\C\alpha\beta}$ and we cut it
down to size {\it exactly} as before, {\it i.e.} we note that if we
introduce another open set $\O\subset\rs$, with $\O\cap\C=\empty$,
and, as in (4) write 
$$\bphi = V(\phi_1,\zeta_1)V(\phi_2,\zeta_2)\cdots
V(\phi_m,\zeta_m)\Omega \in \B_\O \,,\eqno(83)$$
where $\phi_j\in V_{k_j}$, $j=1,\ldots m$, each $\bphi\in\B_\O$
defines a map on $\B_{\C\alpha\beta}$ by 
$$\eta_{\bphi}(\bchi)=(\bphi,\bchi)
=\left\langle\prod_{i=1}^mV(\phi_i,\zeta_i)
\prod_{i=1}^n V(\psi_i,z_i)W_\alpha(\chi_1,u_1)
W_\beta(\chi_2,u_2)\right\rangle\,.\eqno(84)$$
Again $\eta_\bphi$ extends by linearity to a map 
$\V_{\C\alpha\beta}\rightarrow\Cop$ and we consider the space,
$\tV_{\C\alpha\beta}$, of sequences $\bX= (X_1,X_2,\ldots)$,
$X_j\in\V_{\C\alpha\beta}$, for which $\eta_\bphi(X_j)$ converges
uniformly on each of the family of compact sets of the form (7). We
write $\eta_\bphi(\bX)=\lim_{j\rightarrow\infty}\eta_\bphi(X_j)$ and
define the space $\V_{\C\alpha\beta}^\O$ as being composed of the
equivalence classes of such sequences, identifying two sequences
$\bX_1, \bX_2$, if  $\eta_\bphi(\bX_1)=\eta_\bphi(\bX_2)$ for all
$\bphi\in\B_\O$. Using the same arguments as in the proof of Theorem~1
(see Appendix~C), it can be shown that the space
$\V_{\C\alpha\beta}^\O$ is independent of $\C$, provided that the
complement of $\O$ is path-connected; in this case we write
$\V_{\alpha\beta}^\O \equiv \V_{\C\alpha\beta}^\O$. We can define a
family of seminorms for $\V_{\alpha\beta}^\O$ by 
$||\bX ||_\bphi = |\eta_{\bphi}(\bX)|$, where $\bphi$ is an arbitrary
element of $\B_\O$, and the natural topology on $\V_{\alpha\beta}^\O$
is the topology that is induced by this family of seminorms. (This is
to say, that a sequence of states in $\bX_j\in\V_{\alpha\beta}^\O$
converges if and only if $\eta_{\bphi}(\bX_j)$ converges for every 
$\bphi\in\B_\O$.)

So far we have not specified a relationship between the spaces
$\V^\O$, which define the conformal field theory, and the new spaces 
$\V^\O_{\alpha\beta}$, which we have now introduced to define a
representation of it. Such a relation is an essential part of the
definition of a representation; it has to express the idea that the
two spaces define the same relations between combinations of vectors
in the sets $\B_\C$. To do this consider the space of all continuous
linear functionals on $\V^\O$, the {\it dual} space of $\V^\O$, which
we will denote $(\V^\O)^\prime$, and  also the dual,
$(\V_{\alpha\beta}^\O)'$, of $\V_{\alpha\beta}^\O$. It is natural to
consider these dual spaces as topological vector spaces with the weak
topology: for each $f\in (\V^\O)^\prime$, we can consider the
(uncountable) family of seminorms defined by 
$||f||_\Psi\equiv |f(\Psi)|$, where $\Psi$ is an arbitrary 
element of $\V^\O$ (and similarly for 
$(\V_{\alpha\beta}^\O)^\prime$). The weak topology is then the
topology that is induced by this family of seminorms (so that
$f_n\rightarrow f$ if and only if $f_n(\Psi)\rightarrow f(\Psi)$ for
each $\Psi\in\V^\O$).  

Every element of $\bphi\in\B_\O$ defines a continuous linear
functional both on $\V^\O$ and on $\V_{\alpha\beta}^\O$, each of which
we shall denote by $\eta_\bphi$, and the linear span of the set of all
linear functionals that arise in this way is dense in both 
$(\V^\O)^\prime$ and  $(\V_{\alpha\beta}^\O)^\prime$. We therefore 
have a map from a dense subspace of $(\V^\O)^\prime$ to a dense
subspace of $(\V_{\alpha\beta}^\O)^\prime$, and the condition for the 
amplitudes (80) to define a  {\it representation} of the meromorphic
(conformal) field theory whose spaces of states are given by $\V^\O$
is that this map extends to a {\it continuous} map between the dual
spaces, \ie\ that there exists a continuous map
$$
\iota:(\V^\O)^\prime  \rightarrow (\V_{\alpha\beta}^\O)^\prime 
\qquad\hbox{such that}\qquad \iota(\eta_\bphi)=\eta_\bphi\,. \eqno(85)
$$
This in essence says that $\V^\O_{\alpha\beta}$ will not distinguish
limits of linear combinations of $\B_\O$ not distinguished by
$\V^\O$. 

Given a collection of densities $\R$ we can construct (in a similar
way as before for the collection of amplitudes $\A$) spaces
of states $\V_{\alpha}^\O$ and $\V_{\beta}^\O$, on which the vertex
operators of the meromorphic theory are well-defined operators. By the
by now familiar scheme, let us introduce the set $\B_{\C\alpha}$ that
is labelled by finite collections of $\psi_i\in V_{h_i}$, 
$z_i\in\C\subset \rs$, $i = 1,\ldots, n$, $n\in\Nop$
and $z_i\ne z_j$ if $i\ne j$, together with $\chi\in W_\alpha$
and $u\in\C$, $z_i\ne u$, denoted by 
$$\eqalignno{
\bchi &= V(\psi_1,z_1)V(\psi_2,z_2)\cdots V(\psi_n,z_n)
W_\alpha(\chi,u)  \Omega \cr
&\equiv \prod_{i=1}^n V(\psi_i,z_i)W_\alpha(\chi,u)
\Omega\,.&(86)\cr}$$ 
We again immediately identify different $\bchi\in\B_{\C\alpha}$
with the other elements of $\B_{\C\alpha}$ obtained by replacing each
$\psi_j$ in (86) by $\mu_j\psi_j$, $1\leq j\leq n$, $\chi$ by
$\lambda\chi$, where $\lambda, \mu_j\in\Cop$ and 
$\lambda\prod_{j=1}^n\mu_j=1$. We also define $\B_{\C\beta}$
analogously (by replacing $\chi\in W_\alpha$ by $\chi\in W_\beta$). 

We then introduce the vector space $\V_{\C\alpha}$ with basis 
$\B_{\C\alpha}$, and we cut it down to size exactly as before by
considering the map analogous to (84), where now
$\bphi\in\B_{\O\beta}$. The resulting space is denoted by
$\V_{\C\alpha}^{\O\beta}$, and is again independent of $\C$ provided
that the complement of $\O$ is path-connected; in this case we write 
$\V_{\alpha}^{\O\beta} \equiv \V_{\C\alpha}^{\O\beta}$. It also has a
natural topology induced by the seminorms $|\eta_\bphi(\bX)|$, where
now $\bphi\in\B_{\O\beta}$. With respect to this topology, the span of
$\B_{\C\alpha}$ is dense in $\V_{\alpha}^{\O\beta}$. We can 
similarly consider the spaces $\V_{\beta}^{\O\alpha}$ by exchanging
the r\^oles of $W_\alpha$ and $W_\beta$.

For $\psi\in V$, a vertex operator $V(\psi,z)$ can be defined as an
operator 
$V(\psi,z): \V_{\alpha}^{\O\beta}\rightarrow\V_{\alpha}^{\Op\beta}$,
where $z\in\O$ but $z\not\in\Op\subset\O$, by defining its action on
the total subset $\B_{\C\alpha}$, where $\C\cap\O=\empty$ 
$$
V(\psi,z) \bchi = V(\psi,z) V(\psi_1,z_1) \cdots V(\psi_n,z_n) 
W_\alpha(\chi,u) \Omega\,, $$
and $\bchi\in\B_{\C\alpha}$ is as in (86). The image is in 
$\B_{\Cp\alpha}$ for any $\Cp\supset\C$ which contains $z$, and we can
choose $\Cp$ such that $\Cp\cap\Op=\empty$. This then extends by
linearity to a map $\V_{\C\alpha} \rightarrow \V_{\Cp\alpha}$, and we
can show, by analogous arguments as before, that it induces a map 
$\V^{\O\beta}_{\C\alpha} \rightarrow \V^{\Op\beta}_{\Cp\alpha}$.

By the same arguments as before in Section 4, this definition can be
extended to vectors $\Psi$ of the form (34) that span the Fock space
of the meromorphic theory. The actual Fock space $\H^\Op$ (that is
typically a quotient space of the free vector space spanned by
the vectors of the from (34)) is a subspace of $(\V^\O)^\prime$
provided that $\Op\cup\O=\rs$, and if the amplitudes define a
representation, $\iota(\H^\Op)\subset (\V_{\alpha\beta}^\O)^\prime$
because of (85). In this case it is then possible to define vertex
operators  
$V(\Psi,z):\V_{\alpha}^{\O\beta}\rightarrow\V_{\alpha}^{\Op\beta}$  
for arbitrary elements of the Fock space $\H$, and this is what is
usually thought to be the defining property of a representation.
By the same argument the vertex operators are also well-defined
for elements in $\sV_\C$ for suitable $\C$.

There exists an alternative criterion for a set of densities to define
a representation, which is in essence due to Montague\ref{17}, and 
which throws considerable light on the nature of conformal field
theories and their representations. (Indeed, we shall use it to
construct an example of a representation for the $u(1)$-theory below.)

{\bf Theorem 6:} The densities (80) define a representation provided
that, for each open set $\O\subset\rs$ with path-connected complement
and $u_1,u_2\notin\O$, there is a state
$\Sigma_{\alpha\beta}(u_1,u_2;\chi_1,\chi_2)\in\V^\O$ that is
equivalent to $W_\alpha(u_1,\chi_1)W_\beta(u_2,\chi_2)$ in the sense
that the amplitudes of the representation are given by
$\eta_\bphi(\Sigma_{\alpha\beta})$: 
$$\left\langle\prod_{i=1}^m
V(\phi_i,\z_i)W_\alpha(u_1,\chi_1)W_\beta(u_2,\chi_2)\right\rangle
=\left\langle\prod_{i=1}^mV(\phi_i,\z_i)
\Sigma_{\alpha\beta}(u_1,u_2;\chi_1,\chi_2)\right\rangle \,, \eqno(87)
$$
where $\z_i\in\O$.

The proof of this theorem depends on the following:

{\bf Lemma:} There exist sequences $e_i\in\V^\O_\C, f_i\in(\V^\O_\C)'$,
dense in the appropriate topologies, such that $f_j(e_i)=\delta_{ij}$ and 
such that $\sum_{i=1}^\infty e_i f_i(\Psi)$ converges to $\Psi$ for all
$\Psi\in\V^\O_\C$. 

To prove the Lemma, take the $\{e_i\}$ to be formed from the union of
the bases of the eigenspaces $\H_N$ of $L_0$, which we have taken to
be finite-dimensional, taken in order, $N=0,1,2,\ldots$. Using the
projection operators $P_N$ defined by (67), we have that
$\sum_{N=0}^\infty P_N\Psi=\Psi$ and $P_n\Psi$ can be written as a sum
of the $e_i$ which are basis elements of $\H_N$, with coefficients
$f_i(\Psi)$ which depend continuously and linearly on $\Psi$. It is
then clear that $\sum_{i=1}^\infty e_i f_i(\Psi)=\Psi$ and, if
$\eta\in(\V^\O)'$, $\eta = \sum_{i=1}^\infty f_i\eta(e_i)$, showing
that $\{e_i\}$ is dense in $\V^\O$ and $\{f_i\}$ is dense in
$(\V^\O)'$.

{\bf Proof of Theorem 6:}
Assuming we have a continuous map $\iota:(\V^\O)^\prime  \rightarrow
(\V_{\alpha\beta}^\O)^\prime$, let us define
$\Sigma_{\alpha\beta}(u_1,u_2;\chi_1,\chi_2)$ by
$$
\Sigma_{\alpha\beta}(u_1,u_2;\chi_1,\chi_2)=
\sum_ie_i\iota(f_i) 
\Bigl(W_\alpha(u_1,\chi_1)W_\beta(u_2,\chi_2)\Bigr)\,.
\eqno(88)
$$
Then, if $\eta_\bphi= \sum_j\lambda_jf_j$, 
$$
\eqalignno{
\left\langle\prod_{i=1}^m
V(\phi_i,\z_i)W_\alpha(u_1,\chi_1)W_\beta(u_2,\chi_2)\right\rangle
&=
\eta_\bphi\Bigl(\Sigma_{\alpha\beta}(u_1,u_2;\chi_1,\chi_2) \Bigr)\cr
& = \sum_{ji} \lambda_j f_j(e_i)\, \iota(f_i)
   \Bigl( W_\alpha(u_1,\chi_1)W_\beta(u_2,\chi_2) \Bigr) \cr
& = \iota(\eta_\bphi)
    \Bigl( W_\alpha(u_1,\chi_1)W_\beta(u_2,\chi_2) \Bigr) \cr
&=\left\langle\prod_{i=1}^mV(\phi_i,\z_i)
\Sigma_{\alpha\beta}(u_1,u_2;\chi_1,\chi_2)\right\rangle \,,}
$$
and the convergence of (88) can be deduced from this. 

Conversely, suppose that (87) holds; then 
$$
\prod_{i=1}^nV(\psi_i,\zeta_i)
W_\alpha(u_1,\chi_1)W_\beta(u_2,\chi_2) \rightarrow 
\prod_{i=1}^nV(\psi_i,\zeta_i)
\Sigma_{\alpha\beta}(u_1,u_2;\chi_1,\chi_2)
\eqno(89)
$$
defines a continuous map $\V^\O_{\alpha\beta}\rightarrow\V^\O$
(where $\zeta_i,u_1,u_2\notin\O$), and this induces a dual map
$\iota:(\V^\O)'\rightarrow(\V^\O_{\alpha\beta})'$, continuous in the
weak topology, satisfying $\iota(\eta_\bphi)=\eta_\bphi$, \ie\ (85)
holds. 

The map (89) defines an isomorphism of $\V^\O_{\alpha\beta}$ onto
$\V^\O$: it is onto for otherwise its image would define an invariant
subspace of $\V^\O$ and the argument of Proposition 5 shows that this
must be the whole space; and it is an injection because if it maps a
vector $\bX$ to zero, $\eta_\bphi(\bX)$ must vanish for all
$\bphi\in\B_\O$, implying $\bX=0$.

\vskip24pt
\leftline{\bf 9. M\"obius covariance, Fock spaces and the Equivalence
of Representations}
\vskip6pt

We shall now assume that each density in the collection $\R$ is
invariant under the action of the M\"obius transformations, {\it i.e.}
that the amplitudes satisfy
$$\eqalign{
\langle \prod_{i=1}^{n} V(\psi_i,z_i) W_\alpha(\chi_1,u_1) 
W_\beta(\chi_2,u_2)\rangle
= & \langle \prod_{i=1}^{n} V(\psi_i,\gamma(z_i))
W_\alpha(\chi_1,\gamma(u_1)) W_\beta(\chi_2,\gamma(u_2))\rangle \cr
& \qquad \qquad \prod_{l=1}^{2} (\gamma'(u_l))^{r_l} 
\prod_{i=1}^{n} (\gamma'(z_i))^{h_i} \,,} \eqno(90)$$
where $r_l$ are the real numbers which appear in the definition
of the densities, and $h_i$ is the grade of $\psi_i$. 

In this case, we can define operators $U(\gamma)$, mapping
$\V_{\alpha}^{\O\beta}$ to $\V_{\alpha}^{\O_\gamma \beta}$; on the
total subset $\B_{\C\alpha}$, where $\C\cap\O=\empty$, these operators
are defined by   
$$
U(\gamma) \bchi =
V(\psi_1, \gamma(z_1)) \cdots V(\psi_n,\gamma(z_n)) 
W_\alpha(\chi,\gamma(u)) \Omega \, \prod_{i=1}^{n}
(\gamma'(z_i))^{h_i} \, \gamma'(u)^{r_1} \,, \eqno(91)$$
where $\bchi$ is defined as in (86), and $h_i$ is the grade of
$\psi_i$, $i=1,\ldots ,n$. This definition extends by linearity to
operators being defined on $\V_{\C\alpha}$, and by analogous arguments
to those in Section 3, this extends to a well-defined map
$\V_{\alpha}^{\O \beta} \rightarrow \V_{\alpha}^{\O_\gamma \beta}$. 
If we choose two points $z_\infty$ and $z_0$ as before, we can
introduce the M\"obius generators $L^M_0, L^M_{\pm 1}$ which are
well-defined on these spaces. 

We define the Fock space $\H_\alpha^\O\subset \V_\alpha^\O$ to be the
space spanned by finite linear combinations of vectors of the form 
$$
\Phi = V_{n_1}(\psi_1) \cdots V_{n_N}(\psi_N) W_\alpha(\chi,0)
\Omega\,, \eqno(92)
$$
where $\psi_j\in V$, $\chi\in W_\alpha$ and $n_j\in\Zop$, 
$1\leq j\leq N$. Here the modes $V_n(\psi)$ are defined as before in
(32) where the contour encircles the point $0\in\C$, and this still
makes sense since the amplitudes $\R$ are not branched about
$u_i=z_j$. It is clear that $\H_\alpha^\O$ is a dense subspace 
of $\V_\alpha^\O$, and that it is independent of $\O$; where no
ambiguity arises we shall therefore denote it by $\H_\alpha$. By
construction, $W_\alpha\subset \H_\alpha$. We can also define
$W_\beta\subset\H_\beta$ in the same way.

As before it is then possible to extend the amplitudes $\R$ to
amplitudes being defined for $\chi_1\in\H_\alpha$ and
$\chi_2\in\H_\beta$ (rather than $\chi_1\in W_\alpha$ and 
$\chi_2\in W_\beta$), and for the subset of quasiprimary states in 
$\H_\alpha$ and $\H_\beta$ (\ie\ for the states that are annihilated
by $L_1^M$ defined above), the M\"obius properties are analogous to
those in (90).

As in the case of the meromorphic theory we can then define the
equivalence of two representations. Let us suppose that for a given
meromorphic field theory specified by $V$ and $\A=\{f\}$, we have two
collections of densities, one specified by $W_\alpha, W_\beta$ with
the amplitudes given by $\R=\{g\}$, and one specified by 
$\hat{W}_\alpha$, $\hat{W}_\beta$ and $\hat{\R}$. We denote the
corresponding Fock spaces by $\H_\alpha$, $\H_\beta$ in the case of
the former densities, and by $\hat{\H}_\alpha$ and $\hat{\H}_\beta$ in
the case of the latter. We say that the two densities define
{\it equivalent} representations if there exist graded injections
$$
\iota_\alpha : W_\alpha \rightarrow \hat{\H}_\alpha \qquad
\iota_\beta : W_\beta \rightarrow \hat{\H}_\beta \,, \eqno(93)
$$
and
$$
\hat\iota_\alpha : \hat{W}_\alpha \rightarrow \H_\alpha \qquad
\hat\iota_\beta : \hat{W}_\beta \rightarrow \H_\beta \,,\eqno(94)
$$
that map amplitudes to amplitudes. We similarly define two
representations to be {\it conjugate} to one another if instead of
(93) and (94) the amplitudes are mapped to each other under
$$
\iota_\alpha : W_\alpha \rightarrow \hat{\H}_\beta \qquad
\iota_\beta : W_\beta \rightarrow \hat{\H}_\alpha \,, \eqno(93')
$$
and
$$
\hat\iota_\alpha : \hat{W}_\alpha \rightarrow \H_\beta \qquad
\hat\iota_\beta : \hat{W}_\beta \rightarrow \H_\alpha \,.\eqno(94') 
$$

A representation is called {\it highest weight}, if the equivalence 
class of collections of densities contains a representative which has 
the {\it highest weight property}: for each density $g$ and each
choice of $\chi_1\in W_\alpha, \chi_2\in W_\beta$ and 
$\psi_i\in V_{h_i}$, the pole in $(z_i-u_l)$ is bounded by $h_i$.
This definition is slightly more general than the definition which is
often used, in that it is not assumed that the highest weight vectors
transform in any way under the zero modes of the meromorphic fields.

In Section 6, we showed, using the cluster property, that the
meromorphic conformal field theory does not have any proper
ideals. This implies now  

{\bf Proposition 7:} Every non-trivial representation is faithful.

{\bf Proof:} Suppose that $V(\Phi,z)$, where $\Phi\in\sV_\C^\Op$,
$\C\cap\Op=\empty$ and $\C_z\subset\O$, acts trivially on the
representation $\V_{\alpha}^\O$, \ie\ that 
$$
V(\Phi,z) \Psi = 0 \qquad \hbox{for every} \quad
\Psi\in\V_{\alpha}^\O\,. \eqno(95)$$
Then, for any $\psi\in V$ and $\zeta\in\Op$ for which $\zeta+z\in\O$
we have 
$$ 
V\left( V(\psi,\zeta) \Phi,z\right) \Psi =
V(\psi,\zeta + z) V(\Phi,z) \Psi = 0\,, \eqno(96) $$
and thus $V(\psi,\zeta) \Phi$ also acts trivially on 
$\V_{\alpha}^\O$. This implies that the subspace of states in
$\sV_\C^\Op$ that act trivially on $\V_{\alpha}^\O$ is an ideal. Since
there are no non-trivial ideals in $\V_\C^\O$, this implies that 
the representation is faithful.

\vskip24pt
\leftline{\bf 10. An Example of a Representation}
\vskip6pt

Let us now consider the example of the $U(1)$ theory which was first
introduced in Section 5(a). In this section we want to construct a
family of representations for this meromorphic conformal field
theory. 

Let us first define the state
$$
\Psi_n = \int_{a}^{b} dw_1 \cdots \int_{a}^{b} dw_n 
:J(w_1) \cdots J(w_n): \,, \eqno(97) 
$$
where $a,b\in\C\subset\Cop$, the integrals are chosen to lie in $\C$,
and the normal ordering prescription $:\cdot :$ means that all poles
in $w_i-w_j$ for $i\ne j$ are subtracted. We can deduce from the
definition of the amplitudes (48) and (97) that the amplitudes
involving $\Psi_n$ are of the form
$$ \left\langle \Psi_n \prod_{j=1}^{N} J(\z_j) \right\rangle 
= k^n \sum_{i_1, \ldots, i_n \in\{1,\ldots, N\} \atop i_j\ne i_l} 
\prod_{l=1}^{n} {(b-a) \over (a-\z_{i_l}) (b - \z_{i_l})} \;\;\;
\langle \prod_{j\not\in \{i_1, \ldots i_n\}} J(\z_j) \rangle \,, 
\eqno(98)
$$
where $\z_j\in\O\subset\Cop$ and $\C\cap\O=\empty$. By analytic
continuation of (98) we can then calculate the contour integral 
$\oint_{C_a} J(z) dz \Psi_n$, where $C_a$ is a contour in $\C$
encircling $a$ but not $b$, and we find that 
$$ 
\oint_{C_a} J(z) dz \Psi_n = -n k \Psi_{n-1} \,, \eqno(99) $$
and
$$
\oint_{C_a} (z-a)^n J(z) dz \Psi_n = 0 \qquad \hbox{for} \quad
n\geq 1\,, \eqno(100)$$
where the equality holds in $\V^\O$. Similar statements also hold for 
the contour integral around $b$,
$$ 
\oint_{C_b} J(z) dz \Psi_n = n k \Psi_{n-1} \,, \eqno(101) $$
and
$$
\oint_{C_b} (z-b)^n J(z) dz \Psi_n = 0 \qquad \hbox{for} \quad
n\geq 1\,. \eqno(102)$$
Next we define
$$ \Psi_{\alpha}  
= \sum_{n=0}^{\infty} {\alpha^n \over n! \, k^n} \Psi_n 
= : \exp \left( {\alpha \over k} \int_{a}^{b} J(w) dw \right) : \,, 
\eqno(103)$$
where $\alpha$ is any (real) number. This series converges in
$\V^\O$, since for any amplitude of the form
$$
\left\langle \Psi_\alpha \, J(\zeta_1)\cdots J(\zeta_N)
\right\rangle 
$$
only the terms in (103) with $n\leq N$ contribute, as follows from
(98). 

We can use $\Psi$ to define amplitudes as in (87), and in order to
show that these form a representation, it suffices (because of Theorem
6) to demonstrate that the functions so defined have the appropriate
analyticity properties. The only possible obstruction arises from the
singularity for $\zeta_i\rightarrow a$ and $\zeta_i\rightarrow b$, but
it follows from (99-102) that 
$$
J(\zeta) \Psi \sim {-\alpha \over (\zeta-a)} + O(1) \quad 
\hbox{as} \quad \zeta\rightarrow a\,,
$$
and
$$
J(\zeta) \Psi \sim {\alpha \over (\zeta-b)} + O(1) \quad 
\hbox{as} \quad \zeta\rightarrow b\,,
$$
and thus that the singularities are only simple poles. This proves
that the amplitudes defined by (87) give rise to a representation of
the $U(1)$ theory. From the point of view of conventional conformal
field theory, this representation (and its conjugate) is the highest
weight representation that is generated from a state of $U(1)$-charge
$\pm \alpha$. 

It may be worthwhile to point out that we can rescale all amplitudes
of a representation of a meromorphic field theory by 
$$
g \mapsto C (u_1 - u_2)^{2\delta} g \,, \eqno(104)
$$
where $C$ and $\delta$ are fixed constants (that are the same for
all $g$), without actually violating any of the conditions we have
considered so far. (The only effect is that $r_1$ and $r_2$ are
replaced by $\hat{r}_l = r_l - \delta$, $l=1,2$.)  For the
representation of a meromorphic {\it conformal} field theory, the
ambiguity in $\delta$ can however be canonically fixed: since the
meromorphic fields contain the stress-energy field $L$ (whose modes
satisfy the Virasoro algebra $L_n$), we can require that 
$$
L_n = L^M_n \qquad \hbox{for} \quad n=0,\pm 1 \,, \eqno(105)
$$
when acting on $\H_\alpha$. The action of $L_0$ on $\H_\alpha$ is not
modified by (104), but since $\hat{r}_l = r_l - \delta$, the action of
$L_0^M$ is, and (105) therefore fixes the choice of $\delta$ in (104). 

In the above example, in order to obtain a representation of the
meromorphic conformal field theory (with $L^0$ being given by (79)),
we have to modify the amplitudes as in (104) with
$\delta=- \alpha^2/2k$. This can be easily checked using (99-102).

\vskip24pt
{\bf 11. Zhu's Algebra}
\vskip6pt

The description of representations in terms of collections of
densities has a large redundancy in that many different collections
of densities define the same representation. Typically we are only
interested in highest weight representations, and for these we may
restrict our attention to the representatives for which the highest
weight property holds. In this section we want to analyse the
conditions that characterise the corresponding states 
$\Sigma_{\alpha\beta}$; this approach is in essence due to
Zhu\ref{18}.  

Suppose we are given a highest weight representation, {\it i.e.} a
collection of amplitudes that are described in terms of the states
$\Sigma_{\alpha\beta}(u_1,u_2;\chi_1,\chi_2)\in\V^\O$, where
$u_l\not\in\O$. Each such state defines a linear functional on the
Fock space $\H^\Op$, where $\O\cup\Op=\rs$. But, for given $u_1,u_2$,
the states $\Sigma_{\alpha\beta}(u_1,u_2;\chi_1,\chi_2)$, associated
with the various possible representations, satisfy certain linear
conditions: they vanish on a certain subspace
$O_{u_1,u_2}(\H^\Op)$. Thus they define, and are characterised by,
linear functionals on the quotient space
$\H^\Op/O_{u_1,u_2}(\H^\Op)$. This is a crucial realisation, because
it turns out that, in cases of interest, this quotient is
finite-dimensional. Further the quotient has the structure of an
algebra, first identified by Zhu\ref{18}, in terms of which the
equivalence of representations, defined by these linear functionals,
can be characterised.

Let us consider the
case where $u_1=\infty$ and $u_2=-1$, for which we can choose $\O$ and
$\Op$ so that $0\in\O$ and $0\not\in\Op$. We want to characterise the
subspace of $\H=\H^\Op$ on which the linear functional defined by 
$\Sigma_{\alpha\beta}(\infty,-1;\chi_1,\chi_2)$ vanishes
identically. Given $\psi$ and $\chi$ in $\H$, we define the state
$V^{(N)}(\psi) \chi$ in $\H$ by
$$
V^{(N)}(\psi) \chi =
\oint_{0} {d w \over w^{N+1}} 
V\left[ \left( w +1 \right)^{L_0} \psi, w \right] \chi \,,
\eqno(106)
$$
where $N$ is an arbitrary integer, and the contour is a small circle
(with radius less than one) around $w=0$. If $\Sigma_{\alpha\beta}$
has the highest weight property then
$$
\langle \Sigma_{\alpha\beta}(\infty,-1;\chi_1,\chi_2) V^{(N)}(\psi)
\phi \rangle = 0 \qquad \hbox{for $N>0$.} \eqno(107)
$$
This follows directly from the observation that the integrand in (107) 
does not have any poles at $w=-1$ or $w=\infty$. 

Let us denote by $O(\H)$ the subspace of $\H$ that is generated by
states of the form (106) with $N>0$, and define the quotient space
$A(\H) = \H/ O(\H)$. Then it follows that every highest weight
representation defines a linear functional on $A(\H)$. If
two representations induce the same linear functional on $A(\H)$, then
they are actually equivalent representations, and thus the number of
inequivalent representations is always bounded by the dimension of
$A(\H)$. In fact, as we shall show below, the vector space $A(\H)$ has
the structure of an associative algebra, where the product is defined
by (106) with $N=0$. In terms of the states $\Sigma_{\alpha\beta}$ this
product corresponds to 
$$
\langle \Sigma_{\alpha\beta}(\infty,-1;\chi_1,\chi_2) V^{(0)}(\psi)
\phi \rangle = 
\langle \Sigma_{\alpha\beta}(\infty,-1;V_0(\psi)\chi_1,\chi_2) 
\phi \rangle\,. \eqno(108)
$$
One may therefore expect that the different highest weight
representations of the meromorphic conformal field
theory are in one-to-one correspondence with the different
representations of the algebra $A(\H)$, and this is indeed
true\ref{18}. Most conformal field theories of interest have the
property that $A(\H)$ is a finite-dimensional algebra, and there exist
therefore only finitely many inequivalent highest weight
representations of the corresponding meromorphic conformal field
theory; we shall call a meromorphic conformal field theory for which
this is true  {\it finite}. 

In the above discussion the two points, $u_1=\infty$ and $u_2=-1$ were
singled out, but the definition of the quotient space (and the
algebra) is in fact independent of this choice. Let us
consider the M\"obius transformation $\gamma$ which maps
$\infty\mapsto u_1$, $-1\mapsto u_2$ and $0\mapsto 0$ (where 
$u_l\ne 0$); it is explicitly given as 
$$ \gamma(\zeta) = { u_1 u_2 \zeta \over u_2 (\zeta + 1) - u_1}
\hskip8pt \leftrightarrow \hskip8pt
\pmatrix  u_1 u_2  & 0 \\ u_2 & u_2 - u_1 \endpmatrix\,,$$
with inverse
$$\gamma^{-1}(z) = {u_1-u_2 \over u_2} {z \over (z-u_1)}
\hskip8pt \leftrightarrow \hskip8pt
\pmatrix u_1-u_2 & 0 \\ u_2 &  - u_1 u_2\endpmatrix \,.$$
Writing $\psi'= U(\gamma) \psi$ and $\chi'= U(\gamma) \chi$ we then
find (see Appendix~F)
$$
V_{u_1,u_2}^{(N)}(\psi') \chi' = U(\gamma) 
V^{(N)}\left( \psi \right) \chi \,,
\eqno(109)$$
where $V_{u_1,u_2}^{(N)}(\psi)$ is defined by 
$$ 
\eqalign{V_{u_1, u_2}^{(N)}(\psi) \chi = 
\oint_{0} {d w \over w} {u_1 \over (u_1-w)} &
\left( {u_2 \over (u_2-u_1)} {(u_1-w) \over w} \right)^N \cr
& V\left[ \left( {(u_1-w) (u_2-w) \over u_1 u_2} \right)^{L_0} 
e^{{w \over u_1 u_2} L_{1}} \psi, w \right] \chi \,,}\eqno(110)$$
and the contour encloses $w=0$ but not $w=u_l$. We can then also
define $O_{u_1,u_2}(\H)$ to be the space that is generated by states
of the form (110) with $N>0$, and
$A_{u_1,u_2}(\H)=\H/O_{u_1,u_2}(\H)$.  

As $z=0$ is a fixed point of $\gamma$, $U(\gamma):\H\rightarrow \H$,
and because of (109), $U(\gamma): O(\H) \rightarrow O_{u_1,u_2}(\H)$. 
It also follows from (109) with $N=0$ that the product is
covariant, and this implies that the different algebras
$A_{u_1,u_2}(\H)$ for different choices of $u_1$ and $u_2$ are
isomorphic. To establish that the algebra action is well-defined and
associative, it is therefore sufficient to consider the
case corresponding to $u_1=\infty$ and $u_2=-1$. In this
case we write $V^{(0)}(\psi) \chi$ also as $\psi \ast \chi$.

Let us first show that $O_{u_1,u_2}(\H) = O_{u_2,u_1}(\H)$.
Because of the M\"obius covariance it is sufficient to
prove this for the special case, where $u_1=\infty$ and $u_2=-1$. For
this case  we have $V^{(N)}_{\infty,-1}(\psi)=V^{(N)}(\psi)$ as
before, and 
$$
V^{(N)}_{-1,\infty}(\psi) \equiv V^{(N)}_c(\psi) 
= (-1)^N \oint {d w \over w} {1 \over (w+1)} 
\left({w+1 \over w}\right)^N V\left( (w+1)^{L_0} \psi, w \right)\,.
\eqno(111)
$$
The result then follows from the observation that, for $N\geq 1$
$$
{(w+1)^{N-1} \over w^N} =  \sum_{l=1}^{N} 
{N-1 \choose l-1} w^{-l} \,,
$$
and
$$
{1 \over w^N} = \sum_{l=1}^{N} (-1)^{N-l} {N-1 \choose l-1} 
{(w+1)^{l-1} \over w^l} \,.
$$
In particular, it follows from this calculation that (107) also holds 
if $V^{(N)}(\psi)$ is replaced by $V^{(N)}_c(\psi)$. Because of the
definition of $V^{(N)}_c(\psi)$ it is clear that the analogue of (108)
is now  
$$
\langle \Sigma_{\alpha\beta}(\infty,-1;\chi_1,\chi_2) V^{(0)}_c(\psi)
\phi \rangle = 
\langle \Sigma_{\alpha\beta}(\infty,-1;\chi_1,V_0(\psi)\chi_2) 
\phi \rangle\,. \eqno(112)
$$
One should therefore expect that for $N \geq 0$, the action of
$V^{(0)}(\psi)$ and $V^{(N)}_c(\chi)$ commute up to elements of the
form $V^{(M)}_c(\phi)$ where $M>0$ which generate states in $O(\H)$. 
To prove this, it is sufficient to consider the case, where $\psi$ and
$\chi$ are eigenvectors of $L_0$ with eigenvalues $h_\psi$ and
$h_\chi$, respectively; then the commutator
$[V^{(0)}(\psi),V_c^{(N)}(\chi)]$ equals (up to the constant $(-1)^N$
in (111)) 
$$
\eqalign{ & =
\oint\oint_{|\zeta|>|w|} {d\zeta \over \zeta}(\zeta+1)^{h_\psi}
{dw \over w (w+1)} 
\left( {w+1 \over w}\right)^N (w+1)^{h_\chi}
V(\psi,\zeta) V(\chi,w) \cr
& \qquad 
- \oint\oint_{|z|>|\zeta|} {dw \over w (w+1)} 
\left( {w+1 \over w}\right)^N (w+1)^{h_\chi} 
{d\zeta \over \zeta} 
(\zeta+1)^{h_\psi} V(\chi,w) V(\psi,\zeta) \cr
& =
\oint_0 \left\{ \oint_w {d\zeta \over \zeta}
(\zeta+1)^{h_\psi}  V(\psi,\zeta) V(\chi,w) \right\}
{dw \over w (w+1)} \left( {w+1 \over w}\right)^N (w+1)^{h_\chi} }
$$
$$
\eqalign{& =
\sum_{n}
\oint_0 \left\{ \oint_w {d\zeta \over \zeta}
(\zeta+1)^{h_\psi} V(V_n(\psi)\chi,w) (\zeta-w)^{-n-h_\psi} \right\}
{dw \over w (w+1)} \left( {w+1 \over w}\right)^N (w+1)^{h_\chi} 
\cr
& = 
\sum_{n\leq h_\chi} \sum_{l=0}^{n+h_\psi - 1}
(-1)^l {h_\psi \choose l+1-n} \cr
& \qquad \qquad 
\oint_0 {dw \over w (w+1)} \left( {w+1 \over w}\right)^{N+l+1}
(w+1)^{h_\chi-n} V(V_n(\psi)\chi,w) \cr 
& \approx 0 \,,}
$$
where we denote by $\approx$ equality in $\H$ up to states
in $O(\H)$. Because of the M\"obius covariance, it then also follows 
that $[V^{(N)}(\psi),V_c^{(0)}(\chi)]\approx 0$ for $N\geq 0$, 

As $O_{u_1,u_2}(\H) = O_{u_2,u_1}(\H)$, this calculation implies
that the action of $V^{(0)}(\psi)$ is well-defined on the quotient
space. To prove that the action defines an associative algebra, we
observe that in the same way in which $V(\psi,z)$ is uniquely
characterised by the two properties (28a) and (28b), $V^{(0)}(\psi)$
is uniquely determined by the two properties
$$
\eqalign{ V^{(0)}(\psi) \Omega & = \psi \cr
[V^{(0)}(\psi), V_c^{(N)}(\chi)] & \approx 0 \quad 
\hbox{for $N\geq 0$.}}\eqno(113)
$$
Indeed, if $V^{(0)}(\psi_1)$ and $V^{(0)}(\psi_2)$ both satisfy these
properties for the same $\psi$, then
$$
\eqalign{ V^{(0)}(\psi_1) \phi & = V^{(0)}(\psi_1)
V^{(0)}_c(\phi) \Omega \cr
& \approx V^{(0)}_c(\phi) V^{(0)}(\psi_1) \Omega \cr
& = V^{(0)}_c(\phi) \psi\cr 
& = V^{(0)}_c(\phi)  V^{(0)}(\psi_2) \Omega\cr
& \approx V^{(0)}(\psi_2)  V^{(0)}_c(\phi)  \Omega \cr
& = V^{(0)}(\psi_2) \phi \,,}\eqno(114)
$$
where we have used that 
$V^{(0)}_c(\phi) \Omega = \phi$, as follows directly from the
definition of $V^{(0)}_c(\phi)$. This therefore implies that
$V^{(0)}(\psi_1)=V^{(0)}(\psi_2)$ on $A(\H)$. 

It is now immediate that
$$
V^{(0)}(V^{(0)}(\psi) \chi) \approx V^{(0)}(\psi) V^{(0)}(\chi) \,,
\eqno(115)
$$
since both operators commute with $V^{(0)}_c(\phi)$ for arbitrary
$\phi$, and since  
$$ 
V^{(0)}(V^{(0)}(\psi) \chi) \Omega = V^{(0)}(\psi) \chi 
= V^{(0)}(\psi) V^{(0)}(\chi) \Omega\,.\eqno(116)$$
We have thus shown that
$(\psi\ast\chi)\ast\phi\approx\psi\ast(\chi\ast\phi)$. 
Similarly, 
$$
V^{(0)}(V^{(N)}(\psi) \chi) \approx V^{(N)}(\psi) V^{(0)}(\chi) \,,
\eqno(117)$$
and this implies that $\psi_1\ast\phi\approx 0$ if $\psi_1\approx 0$.
This proves that $A(\H)$ forms an associative algebra.

The algebraic structures on  $A_{u_1,u_2}(\H)$ and
$A_{u_2,u_1}(\H)$ are related by
$$
A_{u_1,u_2}(\H) = \Bigl(A_{u_2,u_1}(\H)\Bigr)^{o} \,,
\eqno(118)
$$
where $A^{o}$ is the {\it reverse algebra} as explained in
Appendix~G. Indeed, it follows from (111) and (113) that  
$$
\eqalignno{
V^{(0)}(\psi_1) V^{(0)}(\psi_2) \Omega & = 
V^{(0)}(\psi_1) V^{(0)}_c (\psi_2) \Omega \cr
& = V^{(0)}_c(\psi_2) V^{(0)}(\psi_1) \Omega \cr
& = V^{(0)}_c(\psi_2) V^{(0)}_c (\psi_1) \Omega\,,}
$$
and this implies (118). 

By a similar calculation to the above, we can also deduce that
for $h_\phi>0$
$$\eqalign{
[V^{(0)}(\phi),V^{(0)}(\psi)]
&\approx\sum\limits_{m=0}^{h_\phi+h_\psi-1}
 V^{(0)}(V_{m+1-h_\phi}(\phi)\psi)
\sum\limits_{s=0}^{\min(h_\phi,m)}(-1)^{m+s}\binom
{h_\phi}{s}\cr
&=\sum\limits_{m=0}^{h_\phi-1}
 V^{(0)}(V_{m+1-h_\phi}(\phi)\psi)
\sum\limits_{s=0}^{m}(-1)^{m+s}\binom
{h_\phi}{s}\cr
&=\sum\limits_{m=0}^{h_\phi-1}
\binom{h_\phi-1}{m}
 V^{(0)}(V_{m+1-h_\phi}(\phi)\psi)\cr
&=\oint
 V^{(0)}(V(\phi,\zeta)\psi)(\zeta+1)^{h_\phi-1}d\zeta\,.\cr
} \eqno(119)$$
and
$$\eqalignno{
V^{(1)}(\psi)\Omega&=\oint V(\psi,\zeta)(\zeta+1)^{h_\psi}
{d\zeta\over\zeta^2}\Omega\cr
&=\sum\limits_{n=0}^{h_\psi}
\binom{h_\psi}n V_{n-h_\psi-1}(\psi)\Omega\cr
&=V_{-h_\psi-1}(\psi)\Omega +
h_\psi V_{-h_\psi}(\psi)\Omega\cr
&=(L_{-1}+L_0) V_{-h_\psi}(\psi)\Omega.\cr
}$$
In particular, this implies that $(L_{-1}+L_0)\psi\approx 0$ for
every $\psi\in\H$.

For the Virasoro field $L(z)=V(L,z)$ (119) becomes
$$\eqalignno{
[V^{(0)}(L),V^{(0)}(\psi)]
&=\sum\limits_{m=0}^{1}
\binom{1}{m}  V^{(0)}(V_{m-1}(L)\psi)\cr
&=(L_{-1}+L_0)\psi\approx 0\,,\cr
}$$
which thus implies that $L$ is central in Zhu's algebra.

So far our considerations have been in essence algebraic, in that we
have considered the conditions $\Sigma_{\alpha\beta}$ has to satisfy 
in terms of the linear functional it defines on the Fock space
$\H$. If, however, we wish to reverse this process, and proceed from a
linear functional on $A(\H)$ to a representation of the conformal field
theory, we need to be concerned about the analytic properties of the
resulting amplitudes. To this end, we note that we can perform an
analytic version of the construction as follows. 

In fact, since $\Sigma_{\alpha\beta}$ is indeed an element of 
$\V^\O$ for a suitable $\O$, it actually defines a linear functional
on the whole dual space $\overline{\V_\O}\equiv (\V^\O)^\prime$ (of
which the Fock space is only a dense subspace). Let us denote by
$\overline{O(\V_\O)}$ the completion (in $\overline{\V_\O}$) of the
space that is generated by states of the form (110) with $N>0$ where
now $\psi\in\H$ and $\chi\in\overline{\V_\O}$. By the same
arguments as before, the linear functional associated to  
$\Sigma_{\alpha\beta}$ vanishes then on $\overline{O(\V_\O)}$, and
thus defines a linear functional on the quotient space
$$
A\left(\overline{\V_\O} \right) = \overline{\V_\O} /
\overline{O(\V_\O)} \,. \eqno(120)
$$
It is not difficult to show (see [19] for further details) that a
priori $A\left(\overline{\V_\O} \right)$ is a quotient space of 
$A(\H)$; the main content of Zhu's Theorem [18] is equivalent to:

{\bf Theorem 8 [Zhu's Theorem]}: The two quotient spaces are
isomorphic vector spaces
$$
A\left(\overline{\V_\O} \right) \simeq  A(\H) \,.
$$

{\bf Proof.} It follows from the proof in ref.~[18] that every
non-trivial linear functional on $A(\H)$ (that is defined by 
$\rho(a)=\langle w^\ast, a w \rangle$ where $w$ is an element of a
representation of $A(\H)$, and $w^\ast$ is an element of the
corresponding dual space) defines a non-trivial state
$\Sigma_{\alpha\beta}\in\V^\O$, and therefore a non-trivial element in
the dual space of $A\left(\overline{\V_\O}\right)$. 

The main importance of this result is that it relates the analytic 
properties of correlation functions (which are in essence encoded in
the definition of the space $\V^\O$, {\it etc.}) to the purely
algebraic Fock space $\H$. 

Every linear functional on $A(\overline{\V_\O})$ defines a
highest weight representation of the meromorphic conformal field
theory, and two such functionals define equivalent representations if
they are related by the action of Zhu's algebra as in (108). Because of
Zhu's Theorem there is therefore a one-to-one correspondence between
highest weight representations of the meromorphic conformal field
theory whose Fock space is $\H$, and representations of the algebra
$A(\H)$; this (or something closely related to it) is the form in which
Zhu's Theorem is usually stated. 

Much of the structure of the meromorphic conformal field theory (and
its representations) can be read off from properties of $A(\H)$. For
example, it was shown in ref.~[21] (see also Appendix~G) that if
$A(\H)$ is semisimple, then it is necessarily finite-dimensional, and
therefore there exist only finitely many irreducible representations
of the meromorphic field theory.

\vskip24pt
\leftline{\bf 12. Further Developments}
\vskip6pt

In this paper we have introduced a rigorous approach to conformal
field theory taking the amplitudes of meromorphic fields as a starting
point. We have shown how the paradigm examples of conformal field
theories, \ie\ lattice theories, affine Lie algebra theories and the
Virasoro theory, all fit within this approach. We have shown how to
introduce the concept of a representation of such a meromorphic
conformal field theory by a using a collection of amplitudes which
involve two non-meromorphic fields, so that the amplitudes may be
branched at the corresponding points. We showed how this led naturally
to the introduction of Zhu's algebra and why the condition that this
algebra be finite-dimensional is a critical one in distinguishing
interesting and tractable theories from those that appear to be less
so.

To complete a treatment of the fundamental aspects of conformal field 
theory we should discuss subtheories, coset theories and orbifolds,
all of which can be expressed naturally within the present
approach\ref{19}. It is also clear how to modify the axioms for
theories involving fermions.

It is relatively straightforward to generalise the discussion of
representations to correlation functions involving $N>2$
non-meromorphic fields. The only difference is that in this case,
there are more than two points $u_l$, $l=1,\ldots, N$, at which the
amplitudes are allowed to have branch cuts. The condition that a
collection of such amplitudes defines an $N$-point correlation
function of the meromorphic conformal field theory can then be
described analogously to the case of $N=2$: we consider the vector
space $\V_{\C\balpha}$ (where $\balpha=(\alpha_1,\ldots,\alpha_N)$
denotes the indices of the $N$ vector spaces $W_{\alpha_i}$ that are
associated to the $N$ points $u_1,\ldots, u_N$), whose elements are
finite linear combinations of vectors of the form
$$
V(\psi_1,z_1) \cdots V(\psi_n,z_n) \,
W_{\alpha_1}(\chi_1,u_1) \cdots W_{\alpha_N}(\chi_N,u_N)\Omega \,.
\eqno(121)
$$
We complete this space (and cut it down to size) using the standard
construction with respect to $\B_\O$ and the above set of amplitudes,
and we denote the resulting space by $\V_{\C\balpha}^\O$. The relevant
condition is then that $\B_\O$ induces a continuous map 
$$
(\V^\O)^\prime  \rightarrow (\V_{\balpha}^\O)^\prime \,. \eqno(122)
$$

There also exists a formulation of this condition analogous to (87): a
collection of amplitudes defines an $N$-point correlation function if
there exists a family of states  
$\Sigma_{\balpha}(u_1,\ldots,u_N;\chi_1,\ldots,\chi_N)\in\V^\O_\C$ for
each  $\O,\C$ with $\O\cap\C=\empty$ that is equivalent to 
$W_{\alpha_1}(\chi_1,u_1)\cdots W_{\alpha_N}(\chi_N,u_N)$ in the sense
that  
$$\eqalign{
\Bigl\langle \prod_{i=1}^n
V(\psi_i,z_i) & W_{\alpha_1}(\chi_1,u_1)\cdots W_{\alpha_N}(\chi_N,u_N) 
\Bigr\rangle \cr
& =\left\langle\prod_{i=1}^n V(\psi_i,z_i)
\Sigma_{\balpha}(u_1,\ldots,u_N;\chi_1,\ldots,\chi_N)
\right\rangle \,,} \eqno(123)
$$
where $z_i\in\O$. An argument analogous to Theorem~6 then implies that
(122) is equivalent to (123).

In the case of the two-point correlation functions (or
representations) we introduced a quotient space (120) of the vector
space $\overline{\V_\O} = (\V^\O)^\prime$ that classified the
different highest weight representations. We can now perform an
analogous construction. Let us consider the situation where the $N$
highest weight states are at $u_1=\infty, u_2, \ldots, u_{N}$, and 
define  
$$
V_N^{(M)}(\psi) \chi = \oint_0 {d\zeta \over \zeta^{1+M}}
\left({ \prod_{j=2}^{N}(\zeta - u_j) \over \zeta^{N-2}}
\right)^{h_\psi} 
V(\psi,\zeta) \chi \,, \eqno(124)
$$
where $M$ is an integer, $\psi\in\H$, $\chi\in\overline{\V_\O}$, and
the contour encircles $0$, but does not encircle $u_1,\ldots,u_{N}$. 
We denote by $\overline{O_N(\V_\O)}$ the completion of the
space (in $\overline{\V_\O}$) that is spanned by states of the form  
(124) with $M>0$, and we denote by $A_N(\overline{\V_\O})$ the
quotient space $\overline{\V_\O} / \overline{O_N(\V_\O)}$. (In this
terminology, Zhu's algebra is the space $A_2(\overline{\V_\O})$.)
By the same arguments as in (107) is is easy to see that every state 
$\Sigma_{\balpha}$ that corresponds to an $N$-point function of $N$
highest weights (where the highest weight property is defined as
before) defines a functional on $\overline{\V_\O}$ that vanishes on 
$\overline{O_N(\V_\O)}$, and thus defines a linear functional on 
$A_N(\overline{\V_\O})$.

One can show (see [19] for more details) that the space
$A_N(\overline{\V_\O})$ carries $N$ commuting actions of Zhu's algebra
$A(\H)$ which are naturally associated to the $N$ non-meromorphic
points $u_1,\ldots,u_N$. For example the action corresponding to $u_1$
is given by (124) with $M=0$ and $\psi\in\H$, 
$\psi\circ\phi=V_N^{(0)}(\psi) \phi$. This action is actually
well-defined for $\psi\in A(\H)$ since we have for $L\geq 0$
$$
V_N^{(0)}\left(V^{(L)}(\psi_1)\,\psi_2\right)\,\phi 
\approx V_N^{(L)}(\psi_1) \,V_N^{(0)}(\psi_2) \,\phi\,,\eqno(125)
$$
where we denote by $\approx$ equality in $\overline{\V_\O}$ up to
states in $\overline{O_N(\V_\O)}$. Applying (125) for $L=0$ implies
that the algebra relations of $A(\H)$ are respected, {\it i.e.} that
$$
\left( \psi_1 \ast \psi_2 \right) \circ \phi = \psi_1 \circ
\left( \psi_2 \circ \phi \right) \,,\eqno(126)
$$
where $\ast$ denotes the multiplication of $A(\H)$. Every $N$-point
correlation function determines therefore $N$ representations of Zhu's
algebra, and because of Zhu's Theorem, we can associate $N$
representations of the meromorphic conformal field theory to
it. Conversely, every linear functional on $A_N(\overline{\V_\O})$
defines an $N$-point correlation function, and two functionals define
equivalent such functions if they are related by the actions of
Zhu's algebra; in this way the different $N$-point correlation
functions of the meromorphic conformal field theory are classified by
$A_N(\overline{\V_\O})$.   

There exists also an `algebraic' version of this quotient space, 
$A_N(\H)=\H / O_N(\H)$, where $O_N(\H)$ is generated by the states of
the form (124) where now $\psi$ and $\phi$ are in $\H$. This space is
much more amenable for study, and one may therefore hope that in
analogy to Zhu's Theorem, the two quotient spaces are isomorphic
vector spaces, 
$$
A_N(\overline{\V_\O}) \simeq A_N(\H)\,; 
$$
it would be interesting if this could be established. 

It is also rather straightforward to apply the above techniques
to an analysis of correlation functions on higher genus Riemann
surfaces. Again, it is easy to see that the correlation functions 
on a genus $g$ surface can be described in terms of a state of the
meromorphic conformal field theory on the sphere, in very much the same
way in which $N$-point correlation functions can be defined by (87).
The corresponding state induces a linear functional on
$\overline{\V_\O}$ (or $\H$), and since it vanishes on a certain 
subspace thereof, defines a linear functional on a suitable quotient
space. For the case of the genus $g=1$ surface, the torus, the
corresponding quotient space is very closely related to Zhu's algebra,
and one may expect that similar relations hold more generally. 
\vskip24pt

\leftline{\bf Acknowledgements}
\vskip6pt

We would like to thank Ben Garling, Terry Gannon, Graeme Segal and
Anthony Wassermann for useful conversations. 

M.R.G. is grateful to Jesus College, Cambridge, for a Research
Fellowship and to Harvard University for hospitality during
the tenure of a NATO Fellowship in 1996/97. P.G. is grateful to 
the Mathematische Forschungsinstitut Oberwolfach and the Aspen 
Center for Physics for hospitality in January 1995 and August 
1996, respectively.

\vskip24pt

\leftline{\bf Appendix A: Sequences of holomorphic functions}
\vskip6pt

A sequence of functions $\{f_n\}=f_1, f_2, f_3, \ldots,$ each
defined on a domain $D\subset \Cop$, is said to be {\it uniformly
bounded} on $D$ if there exists a real number $M$ such that
$|f_n(z)|<M$ for all $n$ and $z\in D$. The sequence is said to be 
{\it locally uniformally bounded} on $D$ if, given $z_0\in D$,
$\{f_n\}$ is uniformly bounded on 
$N_\delta(z_0, D) = \{z\in D: |z-z_0|<\delta\}$ for some 
$\delta \equiv \delta(z_0) >0$. 

A sequence of functions $\{f_n\}$ defined on $D$ is said to be 
{\it uniformly convergent} to $f:D\rightarrow \Cop$ if, given
$\epsilon >0$, $\exists~N$ such that $|f_n(z) - f(z)|<\epsilon$ for
all $z\in D$ and $n>N$. We write $f_n\rightarrow f$ uniformly in
$D$. The sequence is said to be  {\it locally uniformly convergent} to
$f:D\rightarrow \Cop$ if, given $z_0\in D$, $f_n\rightarrow f$
uniformally in $N_\delta(z_0,D)$ for some  $\delta\equiv\delta(z_0)>0$.

Clearly, by the Heine-Borel Theorem, a sequence is locally uniformly
bounded on $D$ if and only if it is uniformally bounded on every
compact subset of $D$, and locally uniformly convergent on $D$ if and
only if it is uniformly convergent on every compact subset of $D$. Local
uniformity of the convergence of a sequence of continuous functions
guarantees continuity of the limit and, similarly, analyticity of the
limit of a sequence of analytic functions is guaranteed by local
uniformity of the convergence\ref{22}. 

The following result\ref{23} is of importance in the approach to
conformal field theory developed in this paper:

{\bf Theorem.} If $D$ is an open domain and
$f_n:D\rightarrow\Cop$ is analytic for each $n$, 
$f_n(z)\rightarrow f(z)$ at each $z\in D$, and the sequence $\{f_n\}$
is locally uniformally bounded in $D$, then $f_n\rightarrow f $
locally uniformally in $D$ and $f$ is analytic in $D$. 

[Proof: Again, given $z_0\in D$, 
$\overline{N_\delta(z_0)}=\{z\in\Cop:|z-z_0|\leq\delta\}\subset D$ for
some $\delta>0$ because $D $ is open. Because
$\overline{N_\delta(z_0)}$ is compact, $\exists M$, such that
$|f_n(z)| < M$ for all $n$ and all
$z\in\overline{N_\delta(z_0)}$. First we show that the sequence
$f_n'(z)$ is uniformally bounded on $N_\rho(z_0)$ for all 
$\rho<\delta$. For
$$
 |f_n'(z)| =\left|{1\over 2\pi i}\oint_{\C_\delta(z_0)}
{f_n(\zeta)d\zeta\over(\zeta-z)^2}\right|
\leq { M \delta\over (\delta - \rho)^2} = M_1(\rho), \quad \hbox{say.}$$
Thus for fixed $\rho$, given $\epsilon >0$, $\exists~\delta_1>0$ such
that $|f_n(z) - f_n(z')|<\third \epsilon $ for all values of $n$
provided that $z,z'\in N_\rho(z_0)$ are such that
$|z-z'|<\delta_1(\epsilon)$. Then also $|f(z) - f(z')|<\third \epsilon
$ for $|z-z'|<\delta_1(\epsilon)$. Now we can find a finite number $K$
of points $z_j\in N_\rho(z_0)$, $1\leq j \leq K$,  such that given 
any point in $z\in N_\rho(z_0)$, $|z-z_j|<\delta_1$ for some $j$. Now
each $f_n(z_j)\rightarrow f(z_j)$ and so we can find integers $L_j$
such that $|f(z_j)-f_n(z_j)|<\third\epsilon$ for $n>L_j$. Now if
$n>L=\max_{1\leq j\leq K} \{L_j\}$ and $z\in N_\rho(z_0)$, 
$|f_n(z) -f(z)|\leq |f_n(z) - f_n(z_j)| + |f_n(z_j) - f(z_j)| +
|f(z_j) - f(z)| <\epsilon$, establishing uniform convergence on
$N_\rho(z_0)$ and so local uniform convergence. This is sufficient to
deduce that $f$ is analytic. 

\vskip24pt
\leftline{\bf Appendix B: Completeness of $\V^\O_\C$}
\vskip6pt

We prove that, if a sequence $\chi_j\in\V^\O_\C$, $j = 1,2,\ldots$,
$\eta_\bphi(\chi_j)$ converges on each subset of $\bphi$ of the form
(6) (where the convergence is uniform on (7)), the limit
$\lim_{j\rightarrow\infty}\eta_\bphi(\chi_j)$ necessarily equals
$\eta_\bphi(\chi)$ for some $\chi\in\V_\C^\O$. 

To see this, note that uniform convergence in this sense is 
implied by the (uniform) convergence on a countable collection of such
sets, taken by considering $\epsilon = 1/N$, $N$ a positive integer,
$K$ one of a collection of compact subsets of $\O$ and the $\phi_j$
to be elements of some countable basis. Taken together we obtain in this
way a countable number of conditions for the uniform
convergence. Defining 
$||\Psi||_n=\max_\bphi \max_{\zeta_j} |\eta_\bphi(\Psi)|$,
where $\bphi$ ranges over the first $n$ of these countable conditions
for each $\O$ and the maximum is taken over $\zeta_j$ within (7),
we have a sequence of semi-norms, $||\Psi||_n$, on $\V_\C$, with
$||\Psi||_n\leq ||\Psi||_{n+1}$. Given such a sequence of 
semi-norms, we can define a Cauchy sequence $(\Psi_j)$, 
$\Psi_j\in \V_\C$, by the requirement that
$||\Psi_i-\Psi_j||_n\rightarrow 0$ as $i,j\rightarrow \infty$ for each
fixed $n$. This requirement is equivalent to uniform convergence on
each set of the form (6). Moreover, the space $\V^\O_\C$ is
obtained by adding in the limits of these Cauchy sequences
(identifying points zero distance apart with respect to all of the
semi-norms). This space is necessarily complete because if
$\chi_j\in\V^\O_\C$ is Cauchy, {\it i.e.}
$||\chi_i-\chi_j||_n\rightarrow 0$ as $i,j\rightarrow \infty$ for each
fixed $n$, and $\Psi_i^m\rightarrow \chi_i$ as $m\rightarrow\infty$,
$\Psi_i^m\in\V_\C$, then selecting $I_N$ so that
$||\chi_i-\chi_j||_N<1/3N$ for $i,j\geq I_N$, and $I_{N+1}\geq I_N$,
we can find an integer $m_N$ such that
$||\psi_{I_N}^{m_N}-\chi_{I_N}||_N<1/3N$, and if
$\psi_N=\psi_{I_N}^{m_N}$, 
$$\eqalignno{
||\psi_M-\psi_N||_p &\leq ||\psi_M-\chi_{I_M}||_p +
||\chi_{I_M}-\chi_{I_N}||_p +||\chi_{I_N} - \psi_N||_p\cr
&\leq ||\psi_M-\chi_{I_M}||_M + ||\chi_{I_M}-\chi_{I_N}||_N
+||\chi_{I_N} - \psi_N||_N\cr
&\leq 1/N\cr}
$$
provided that $M\geq N\geq p$, implying that $\psi_M$ is Cauchy. It is
easy to see that its limit is the limit of $\chi_j$, showing that
$\V_\C^\O$ is complete. The completeness of this space is equivalent
to the condition (11).

\vskip24pt
\leftline{\bf Appendix C: Proof that $\V_\C^\O$ is independent of $\C$
(Theorem 1)} 
\vskip6pt

{\bf Proof:} To prove that $\tV^\O_\C$ is independent of $\C$,
first note that we may identify a vector $\chi\in\tV^\O_\C$ with
$\bpsi = \prod_{i=1}^n V(\psi_i,z_i)\Omega$ (where $z_i\notin\O$ for
$1\leq i \leq n$ but it is not necessarily the case that $z_i\in\C$
for each $i$) if $\eta_\bphi(\chi) = \eta_\bphi(\bpsi)$, for all 
$\bphi\in\B_\O$, \ie\ the value of $\eta_\bphi(\chi)$ is given
by (5) for all $\bphi\in\B_\O$. Consider then the set $\Q$ of
values of $\bz = (z_1,z_2,\ldots,z_n)$ for which 
$\bpsi(\bz) =\prod_{i=1}^n V(\psi_i,z_i)\,\Omega$ is a member of
$\tV^\O_\C$. Then $\Dp\subset\Q\subset\D$, where 
$\Dp = \{\bz : z_i,z_j \in \C, z_i\ne z_j, 1\leq i < j \leq n \}$ and
$\D=\{\bz : z_i, z_j \in \O^c, z_i\ne z_j, 1\leq i < j \leq n \}$,
where $\O^c$ is the complement of $\O$ in $\rs$. We shall show that
$\Q=\D$.   

If $\bz_b$ is in $\D^o$, the interior of $\D$, but not in $\Q$, choose
a point $\bz_a\in\Dp\subset\Q$ and join it to $\bz_b$ by a path $C$
inside $\D^o$, $\{\bz(t): 0\leq t\leq 1\}$ with $\bz(0)=\bz^a$ and 
$\bz(1)=\bz^b$.  (There is such a point $\bz^a$ because
$\C^o\ne\empty$; the path $C$ exists because the interior of $\O^c$ is
connected, from which it follows that $\D^o$ is.)  Let $t_c$ be the
supremum of the values of $t_0$ for which 
$\{\bz(t): 0\leq t\leq t_0\} \subset \Q^o$, the interior of $\Q$, and
let $\bz^c = \bz(t_c)$.  Then $\bz^c = (z_1^c,z_2^c,\ldots,z_n^c)$ is
inside the open set $\D^o$ and so we can find a neighbourhood of the
form $N_1 =\{\bz : |\bz - \bz^c| < 4\delta \}$ which is contained
inside $\D^o$. Let $\bz^d$, $\bz^e$ be points each distant less than
$\delta$ from $\bz^c$, with $\bz^d$ outside $\Q$ and $\bz^e$ inside
$\Q^o$.  (There must be a point $\bz^d$ outside $\Q$ in every
neighbourhood of $\bz^c$.)  Then the set 
$N_2 = \{\bz^e + (\bz^d-\bz^e)\omega : |\omega| < 1\}$ is inside $N_1$
but contains points outside $\Q$.  We shall show that $N_2\subset\Q$ 
establishing a contradiction to the assumption that there is a point
$\bz_b$ in $\D^o$ but not in $\Q$, so that we must have
$\D^o\subset\Q\subset\D$. 

The circle $\{\bz^e + (\bz^d-\bz^e)\omega : |\omega| < \epsilon\}$ is
inside $\Q$ for some $\epsilon$ in the range $0 < \epsilon < 1$. Now,
we can form the integral 
$$\chi = \int_S \bpsi(\bz) \mu(\bz) d^r\bz$$ 
of $\bpsi(\bz)$ over any compact $r$-dimensional sub-manifold
$S\subset \Q$, with continuous weight function $\mu(\bz)$, to obtain
an element $\chi\in\tV^\O_\C$, because the approximating sums to the
integral will have the necessary uniform convergence property. So the
Taylor coefficients 
$$\bpsi_N=\int_{|\omega|=\epsilon}\bpsi(\bz^e +
(\bz^d-\bz^e)\omega)\omega^{-N-1}d \omega \in\tV^\O_\C$$ 
and, since the $\sum_{N=0}^\infty \eta_\bphi(\bpsi_N)\omega^N$
converges to $\eta_\bphi(\bpsi(\bz^e + (\bz^d-\bz^e)\omega))$ for
$|\omega|<1$ and all $\bphi\in\B_\O$, we deduce that $N_2\subset\Q$,
hence proving that $\D^o\subset\Q\subset\D$. Finally, if $\bz_j$ is a
sequence of points in $\Q$ convergent to $\bz_0\in\D$, it is
straightforward to see that $\bpsi(\bz_i)$ will converge to
$\bpsi(\bz_0)$, so that $\Q$ is closed in $\D$, and so must equal
$\D$.

\vfil\eject
\leftline{\bf Appendix D. M\"obius Transformation of Vertices}
\vskip6pt

By virtue of the  Uniqueness Theorem we can establish the
transformation properties 
$$\eqalignno{
e^{\lambda L_{-1}}V(\Psi,z)e^{-\lambda L_{-1}} &=
V(\Psi,z+\lambda)&(\hbox{D}.1)\cr 
e^{\lambda L_0}V(\Psi,z)e^{-\lambda L_0} &= e^{\lambda
h}V(\Psi,e^\lambda z)&(\hbox{D}.2)\cr 
e^{\lambda L_1}V(\Psi,z)e^{-\lambda L_1}&=
(1-\lambda z)^{-2h}V\left(\exp(\lambda(1-\lambda
z)L_1)\Psi,z/(1-\lambda z)\right) \,, &(\hbox{D}.3)\cr 
}$$
where we have used the relation
$$ \pmatrix 1&0\\ -\lambda&1\endpmatrix
\pmatrix 1&z\\ 0&1\endpmatrix =
\pmatrix 1&{z\over 1-\lambda z}\\ 0&1\endpmatrix
\pmatrix 1&0\\ {-\lambda(1-\lambda z)}&1\endpmatrix
\pmatrix 1\over 1- \lambda z&0\\ 0&{1-\lambda z}\endpmatrix\,.
\eqno(\hbox{D}.4)$$
{}From these if follows that
$$\eqalignno{
\langle V(\Phi,z)\rangle &=0 &(\hbox{D}.5)\cr
\langle V(\Phi,z)V(\Psi,\zeta)\rangle & = {\varphi(\Phi,\Psi)\over
(z-\zeta)^{h_\Phi+h_\Psi}}\,, &(\hbox{D}.6)\cr}$$
where $L_0\Phi=h_\Phi\Phi$, $L_0\Psi=h_\Psi\Psi$ and 
$h_\Phi\ne 0$. The bilinear form $\varphi(\Phi,\Psi)$, defined as
being the constant of proportionality in (D.6), has the symmetry
property  
$$\varphi(\Phi,\Psi)=(-1)^{h_\Phi+h_\Psi}\varphi(\Psi,\Phi)\,.
\eqno(\hbox{D}.7)$$
If, in addition, $L_1\Phi=L_1\Psi=0$, it follows from (D.3) applied to
(D.6) that $\varphi(\Psi,\Phi) = 0$ unless $h_\Phi=h_\Psi$.  

It follows from (D.3) that, if $\Psi\in\H_1$, 
$$\langle V(\Psi,z)\rangle =  
(1-\lambda z)^{-2}\langle V\left(\Psi,z/(1-\lambda z)\right)\rangle + 
{\lambda \over (1-\lambda z)}
\langle V\left(L_1\Psi,z/(1-\lambda z)\right)\rangle\,,
\eqno(\hbox{D}.8)$$ 
and from (D.5) that both the left hand side and the first term on the
right hand side of (D.8) vanish, implying that the second term on the
right hand side also vanishes. But $L_1\Psi\in\H_0$ and, if we assume
cluster decomposition, so that the vacuum is unique,
$L_1\Psi=\kappa\Omega$ for some $\kappa\in\Cop$. We deduce that
$\kappa = \langle V(L_1\Psi,\zeta)\rangle = 0$, so that $\Psi$ is
quasi-primary, {\it i.e.} $\H_1=\H_1^Q$. 

We can show inductively that $\H_h$ is the direct sum of spaces
$L_{-1}^n\H_{h-n}^Q$ where $0\leq n<h$, {\it i.e.} $\H$ is composed
of quasi-primary states and their descendants  under the action of
$L_{-1}$. Given $\Psi\in\H_h$, we can find $\Phi\in L_{-1}H_{h-1}$ 
such that $L_1(\Psi-\Phi)=0$; then $\Psi$ is the sum of the
quasi-primary state $\Psi-\Phi$ and $\Phi$ which, by an inductive
hypothesis, is the sum of descendants of quasi-primary states. To find
$\Phi$, note 
$$L_1(\Psi + \sum_{n=1}^h a_n L_{-1}^n L_1^n \Psi) = 
\sum_{n=1}^h (a_{n-1} + a_n(2nh + n(n+1)) L_{-1}^{n-1}L_1^n\Psi\,,$$
where $a_0=1$. So choosing $a_n = - a_{n-1}/(2nh + n(n+1))$, 
$1\leq n\leq h$, we have $L_1(\Psi -\Phi)=0$ for 
$\Phi = - \sum_{n=1}^h a_nL_{-1}^nL_1^n \Psi\in L_{-1}H_{h-1}$
establishing the result.

\vskip24pt
\leftline{\bf Appendix E: Proof that the extended amplitudes $\hat\A$
satisfy the axioms}
\vskip6pt

An alternative description of the extended amplitudes can be given as
follows: we define the amplitudes involving vectors in $\hat{V}$
recursively (the recursion being on the number of times $L$ appears in
an amplitude) by
$$
\langle L(w) \rangle = 0 \eqno(\hbox{E}.1)
$$
$$
\eqalign{
\langle L(w) \prod_{i=1}^{n} V(\psi_i, z_i) \rangle & = 
\sum_{l=1}^n {c_{\psi_l} \over (w-z_l)^4} \,
\langle \prod_{i\neq l} V(\psi_i,z_i) \rangle \cr
& \quad + \sum_{l=1}^{n} {h_l \over (w-z_l)^2} \, \langle 
\prod_{i=1}^{n} V(\psi_i, z_i) \rangle \cr
& \quad + \sum_{l=1}^{n} {1 \over (w-z_l)} \; {d \over dz_l}\,\langle
V(\psi_1,z_1) \cdots V(\psi_l, z_l) \cdots V(\psi_n, z_n) \rangle}
\eqno(\hbox{E}.2)
$$
and
$$
\eqalign{
\langle L(w) \prod_{j=1}^{m} L(w_j)
\prod_{i=1}^{n} V(\psi_i, z_i) \rangle & = 
\sum_{k=1}^{m} {2 \over (w-w_k)^2} \, \langle \prod_{j=1}^{m} L(w_j)
\prod_{i=1}^{n} V(\psi_i, z_i) \rangle \cr
& \quad + \sum_{l=1}^{n} {h_l \over (w-z_l)^2} \, \langle 
\prod_{j=1}^{m} L(w_j) \prod_{i=1}^{n} V(\psi_i, z_i) \rangle \cr 
& \quad + \sum_{k=1}^{m} {1 \over (w-w_k)} \; {d \over dw_k}\,\langle 
\prod_{j=1}^{m} L(w_j) \prod_{i=1}^{n} V(\psi_i, z_i) \rangle
\cr
& \quad + \sum_{l=1}^{n} {1 \over (w-z_l)} \; {d \over dz_l}\,\langle 
\prod_{j=1}^{m} L(w_j) \prod_{i=1}^{n} V(\psi_i, z_i) \rangle
\cr
& \quad + \sum_{k=1}^{m} {c/2 \over (w-w_k)^4} \, \langle 
\prod_{j\neq k} L(w_j) \prod_{i=1}^{n} V(\psi_i, z_i) \rangle \cr
& \quad + \sum_{l=1}^{n} {c_{\psi_l} / 2 \over (w-z_l)^4} \,
\langle \prod_{j=1}^{m} L(w_j) \prod_{i\neq l} V(\psi_i,z_i)
\rangle\,. } 
\eqno(\hbox{E}.3)
$$
Here $h_i$ is the grade of the vector $\psi_i\in V$, $c$ is an
arbitrary (real) number, and $c_\psi$ is zero unless $\psi$ is of
grade two. It is not difficult to see that the functions defined by 
(E.1)--(E.3) agree with those defined in the main part of the text:
for a given set of fields, the difference between the two amplitudes
does not have any poles in $w_j$, and therefore is constant as a
function of $w_j$; this constant is easily determined to be zero.

The diagrammatical description of the amplitudes immediately implies
that the amplitudes are local. We shall now use the formulae
(E.1)--(E.3) to prove that they are also M\"obius covariant. 
The M\"obius group is generated by translations, scalings and the
inversion $z\mapsto 1/z$. It is immediate from the above formulae that
the amplitudes (with the grade of $L$ being $2$) are covariant under
translations and scalings, and we therefore only have to check the
covariance under the inversion $z\mapsto 1/z$. First, we calculate 
(setting for the moment $c_\psi=0$ for all $\psi$ of grade two)
$$ 
\eqalign{
\langle L(1/w) & \prod_{i=1}^{n} V(\psi_i, 1/z_i) \rangle = 
\sum_{l=1}^{n} {h_l \over (1/w-1/z_l)^2} \, \langle \prod_{i=1}^{n} 
V(\psi_i, 1/z_i) \rangle \cr
& + \sum_{l=1}^{n} {1 \over (1/w-1/z_l)} \left. \;
{d \over d \tilde{z}_l} \, \langle V(\psi_1, 1/z_1) \cdots V(\psi_l,
\tilde{z}_l) \cdots V(\psi_n, 1/z_n) \rangle 
\right|_{\tilde{z}_l=1/z_{l}}\,.} \eqno(\hbox{E}.4) 
$$ 
Using the M\"obius covariance of the original amplitudes, we find
$$ 
\eqalignno{
{d \over d \tilde{z}_l} \langle V(\psi_1, 1/z_1) & \cdots  
 V(\psi_l, \tilde{z}_l) \cdots V(\psi_n, 1/z_n) \rangle 
\Bigr|_{\tilde{z}_l= 1/z_{l}}
= -z_l^2 {d \over dz_{l}} \langle \prod_{i=1}^{n} 
V(\psi_i, 1/z_i) \rangle \cr
& = -z_l^2 \prod_{i\neq l} \left( {-1 \over z_i^2}
\right)^{-h_i} {d \over dz_{l}} \left[ \left( {-1 \over z_l^2}
\right)^{-h_l} \; \langle \prod_{i=1}^{n} V(\psi_i, z_i) \rangle
\right] \cr
&=  -z_l^2 \prod_{i=1}^{n} \left( {-1 \over z_i^2} \right)^{-h_i}
{d \over dz_l} \, \langle V(\psi_1, z_1) \cdots
V(\psi_l, z_l) \cdots V(\psi_n, z_n) \rangle \cr
&  \qquad 
-z_l^2 \prod_{i=1}^{n} \left( {-1 \over z_i^2} \right)^{-h_i}
(- h_l)\, {2 \over z_l^3} \left( - {1 \over z_l^2} \right)^{-1}
\langle \prod_{i=1}^{n} V(\psi_i, z_i) \rangle \cr
&= \prod_{i=1}^{n} \left( {-1 \over z_i^2} \right)^{-h_i}
\left[ -z_l^2 {d \over dz_l} \, \langle V(\psi_1, z_1) \cdots
V(\psi_l, z_l) \cdots V(\psi_n, z_n) \rangle \right. \cr
&  \qquad
\left. - 2\; h_l \; z_l\; \langle \prod_{i=1}^{n} V(\psi_i, z_i)
\rangle\right]\,.}
$$
Inserting this formula in the above expression, we get
$$ 
\eqalignno{
\langle & L(1/w) \prod_{i=1}^{n} V(\psi_i, 1/z_i) \rangle \cr
& = \left( {-1 \over w^2} \right)^{-2} \;
\prod_{i=1}^{n} \left( {-1 \over z_i^2} \right)^{-h_i} 
\left\{ \sum_{l=1}^{n}\, h_l \left[ {z_l^2 \over w^2 \, (w-z_l)^2} +
{2 z_l^2 \over w^3 \, (w-z_l)} \right] 
\langle \prod_{i=1}^{n} V(\psi_i, z_i) \rangle \right. \cr
& \qquad \left. + \sum_{l=1}^{n} {z_l^3 \over w^3 \, (w-z_l)} \;
{d \over dz_l} \, \langle V(\psi_1, z_1) \cdots
V(\psi_l, z_l) \cdots V(\psi_n, z_n) \rangle \right\}}
$$
$$ 
\eqalignno{
& = \left( {-1 \over w^2} \right)^{-2} \;
\prod_{i=1}^{n} \left( {-1 \over z_i^2} \right)^{-h_i} \left\{
\sum_{l=1}^{n}\, h_l \, {3 \, z_l^2 \, w - 2 \, z_l^3 \over w^3 \,
(w-z_l)^2}  \; \langle \prod_{i=1}^{n} V(\psi_i, z_i) \rangle \right.
\cr
& \qquad \left. + \sum_{l=1}^{n} {z_l^3 \over w^3 \, (w-z_l)} \;
{d \over dz_l} \, \langle V(\psi_1, z_1) \cdots
V(\psi_l, z_l) \cdots V(\psi_n, z_n) \rangle \right\} \,.}
$$
It remains to show that the expression in brackets actually agrees
with  (E.4). To prove this, we observe, that because of M\"obius 
invariance of the amplitudes we have
$$ \sum_{l=1}^{n}\; {d \over dz_l} \, \langle V(\psi_1, z_1) \cdots
V(\psi_l, z_l) \cdots V(\psi_n, z_n) \rangle  =0\,,$$
$$ \sum_{l=1}^{n}\, h_l\, \langle \prod_{i=1}^{n} V(\psi_i, z_i) \rangle
+ \sum_{l=1}^{n} z_l \; {d \over dz_l} \, \langle V(\psi_1, z_1) \cdots
V(\psi_l, z_l) \cdots V(\psi_n, z_n) \rangle  =0$$
and
$$ \sum_{l=1}^{n} 2 \; h_l \; z_l \, \langle \prod_{i=1}^{n} 
V(\psi_i,z_i) \rangle + \sum_{l=1}^{n} z_l^2 \; {d \over dz_l}\,
\langle V(\psi_1,z_1) \cdots V(\psi_l, z_l) \cdots V(\psi_n, z_n)
\rangle  =0\,. 
$$ 
The claim then follows from the observation that
$$ 
\eqalignno{
& \qquad \sum_{l=1}^{n} h_l \left[ 
{1 \over (w-z_l)^2}  - {3 \, z_l^2 \, w - 2 \, z_l^3 \over w^3 \,
(w-z_l)^2} \right]  
\langle \prod_{i=1}^{n} V(\psi_i, z_i) \rangle \cr
& \qquad  + \sum_{l=1}^{n} \left[ {1 \over (w-z_l)} 
- {z_l^3 \over w^3 \, (w-z_l)}  \right]
{d \over dz_l} \, \langle V(\psi_1, z_1) \cdots
V(\psi_l, z_l) \cdots V(\psi_n, z_n) \rangle \cr
&= {1 \over w^3} \left\{ w^2 \sum_{l=1}^{n}\; {d \over dz_l} \, \langle 
\prod_{i}^{n} V(\psi_i,z_i) \rangle \right. \cr
& \qquad + w \left[\sum_{l=1}^{n} h_l \;
\langle \prod_{i=1}^{n} V(\psi_i, z_i) \rangle
+ \sum_{l=1}^{n} z_l \; {d \over dz_l} \, \langle \prod_{i}^{n}
V(\psi_i,z_i) \rangle \right] \cr
& \qquad + \left. 
\sum_{l=1}^{n} 2 \; h_l \; z_l \;  \langle \prod_{i=1}^{n} V(\psi_i,
z_i) \rangle 
+ \sum_{l=1}^{n} z_l^2 \; {d \over dz_l} \, \langle \prod_{i=1}^{n} 
V(\psi_i, z_i) \rangle \right\} \cr
&= 0\,.}
$$
We have thus shown that the functions of the form (E.2) (for 
$c_\psi=0$) have the correct transformation property under M\"obius
transformations. This implies, as $L$ is quasiprimary, that the
functions of the form (E.3) have the right transformation property for
$c=0$. However, the sum involving the $c$-terms has also (on its own)
the right transformation property, and thus the above  functions have.
This completes the proof. 

Finally, we want to show that the amplitudes $\hat\A$ have the cluster
property provided the amplitudes $\A$ do. We want to prove the
cluster property by induction on the number $N_L$ of $L$-fields in the
extended amplitudes. If $N_L=0$, then the result follows from the
assumption about the original amplitudes. Let us therefore assume that
the result has been proven for $N_L=N$, and consider the amplitudes
with $N_L=N+1$. For a given amplitude, we subdivide the fields into
two groups, and we consider the limit, where the parameters $z_i$ of
one group are scaled to zero, whereas the parameters $\z_j$ of the
other group are kept fixed. Because of the M\"obius covariance, we may
assume that the group whose parameters $z_i$ are scaled to zero
contain at least one $L$-field, $L(z_1)$, say, and we can use (E.2) 
(or (E.3)) to rewrite the amplitudes involving $L(z_1)$ in terms of
amplitudes which do not involve $L(z_1)$ and which have $N_L\leq N$.
It then follows from (E.2) (or (E.3)) together with the induction
hypothesis that the terms involving $(z_1-\z_j)^{-l}$ (where $l=1,2$
or $l=4$) are not of leading order in the limit where the $z_i$ are
scaled to zero, whereas all terms with $(z_1-z_i)^{-l}$ are. This
implies, again by the induction hypothesis, that the amplitudes
satisfy the cluster property for $N_L=N+1$, and the result follows by
induction. 

\vskip24pt
\leftline{\bf Appendix F: M\"obius transformation of Zhu's modes}
\vskip6pt

We want to prove formula (110) in this Appendix. We have to show
that
$$\eqalignno{
V_{u_1,u_2}^{(N)} (\psi) & =
U(\gamma) V^{(N)} (U(\gamma)^{-1} \psi) U(\gamma)^{-1} \cr
& = U(\gamma) \oint_{0} 
V \left[ (\zeta+1)^{L_0} U(\gamma)^{-1} \psi, \zeta \right]
{d \zeta \over \zeta^{N+1}} U(\gamma)^{-1} \cr
& =  \oint_{0}  U(\gamma) 
V \left[ (\zeta+1)^{L_0} U(\gamma)^{-1} \psi, \zeta\right]
U(\gamma)^{-1} {d \zeta \over \zeta^{N+1}} \,.}$$
We therefore have to find an expression for the transformed vertex
operator. By the uniqueness theorem, it is sufficient to evaluate 
the expression on the vacuum; then we find
$$ \eqalignno{
U(\gamma) V \left[ (\zeta+1)^{L_0} U(\gamma)^{-1} \psi, \zeta \right]
U(\gamma)^{-1}  \Omega & =
U(\gamma) e^{\zeta L_{-1}} (\zeta+1)^{L_0} U(\gamma)^{-1} \psi \,.}$$
To calculate the product of the M\"obius transformations, we
write them in terms of $2\times 2$ matrices, determine their product
and rewrite the resulting matrix in terms of the generators
$L_{0}, L_{\pm 1}$. After a slightly lengthy calculation we
then find
$$  U(\gamma) e^{\zeta L_{-1}} (\zeta+1)^{L_0} U(\gamma)^{-1} \psi  =
\hskip150pt $$
$$ \hskip60pt
V \left[ \left( {\zeta +1 \over 
( 1 - {u_2 \zeta \over (u_1-u_2)} )^2} \right)^{L_0} 
\exp \left( {\zeta \over u_2 \zeta + (u_2 -u_1)} L_{1} \right) \psi ,
{u_1 u_2 \zeta \over u_2 \zeta + (u_2-u_1)  } \right] \Omega \,.$$
In the integral for $V_{u_1,u_2}^{(N)}$ we then change
variables to
$$ w = {u_1 u_2 \zeta \over u_2 \zeta + (u_2-u_1)} = 
\gamma(\zeta) \,;$$
in terms of $w$ the relevant expressions become
$$ 1 + \zeta = {u_1 (w-u_2) \over u_2 (w-u_1)} \hskip20pt
1 - {u_2 \zeta \over (u_1-u_2)} =  {u_1 \over (u_1-w)}  \hskip20pt
d \zeta = {u_1 (u_2-u_1) \over u_2} {d w \over  (w-u_1)^2}  \,.$$
Putting everything together, we then obtain formula (110).

\vskip24pt
\leftline{\bf Appendix G: Rings and Algebras}
\vskip6pt

In this Appendix we review various concepts in algebra; the treatment
follows closely the book [24].

We restrict attention to rings, $R$ which have a unit element, 
$1\in R$. An algebra, $A$, over a field $F$, is a ring which is also
vector space over $F$ in such a way that the structures are compatible
[$\lambda (xy) = (\lambda x) y$, $\lambda\in F$, $x, y\in R$]. The
dimension of $A$ is its dimension as a vector space. We shall in
general consider complex algebras, \ie\ algebras over
$\Cop$. Since $1\in A$ we have $F\subset A$. 

A (left) module for a ring $R$ is an additive group $M$ with a map
$R\times M \rightarrow M$, compatible with the structure of $R$ [\ie\
$(rs)m= r(sm), (r+s)m=rm+sm,\quad r,s\in R, m\in M$]. A module $M$ for
an algebra $A$, viewed as a ring, is necessarily a vector space over
$F$ (because $F\subset A$) and provides a representation of $A$ as 
an algebra in terms of endomorphisms of the vector space $M$.

$R$ provides a module for itself, the adjoint module. A submodule $N$
of a module $M$ for $R$ is an additive subgroup of $M$ such that
$rN\subset N$ for all $r\in R$. A simple or irreducible module is one
which has no proper submodules. A (left) ideal $J$ of $R$ is a
submodule of the adjoint module, {\it i.e.} an additive subgroup
$J\subset R$ such that $rj\in I$ for all $r\in R, j\in J$. The direct
sum $M_1\oplus M_2$ of the $R$ modules $M_1, M_2$ is the additive
group $M_1\oplus M_2$ with 
$r(m_1,m_2)=(rm_1,rm_2), r\in R, m_1\in M_1, m_2\in M_2$. The direct
sum of a (possibly infinite) set  $M_i,i\in I$, of $R$ modules
consisted of elements $(m_i, i\in I)$, with all but finitely many
$m_i=0$. The module $M$ is decomposable if it can be written as the
direct sum of two non-zero modules and completely reducible if it can
be written as the direct sum of a (possibly infinite) sum of
irreducible modules. 

A representation of an algebra $A$ is irreducible if it is irreducible
as a module of the ring $A$.

An ideal $J$ is maximal in $R$ if $K\supset J$ is another ideal in $R$
then $K=R$. If $M$ is an irreducible module for the ring $R$, then
$M\cong R/J$ for some maximal ideal $J\subset R$. [Take 
$m\in M, m\ne 0$ and consider $Rm\subset M$. This is a submodule, so
$Rm=M$. The kernel of the map $r\mapsto rm$ is an ideal, 
$J\subset R$. So $M\cong R/J$. If $J\subset K\subset R$ and $K$ is an
ideal then $K/J$ defines a submodule of $M$, so that $K/J=M$ and
$K=R$, \ie\ J is maximal.] Thus an irreducible representation of a
finite-dimensional algebra $A$ is necessarily finite-dimensional. 

The coadjoint representation $A'$ of an algebra $A$ is defined on the
dual vector space to $A$ consisting of linear maps 
$\rho : A\rightarrow F$ with $(r\rho)(s)=\rho(sr)$. If $M$ is an
$n$-dimensional irreducible  representation of $A$ and $d_{ij}(r)$ the
corresponding representation matrices, the $n$ elements $d_{ij}(r)$,
$1\leq j\leq n$, $i$ fixed, define an $n$-dimensional invariant
subspace $A'$ corresponding to a representation equivalent to $M$. So
the sum of the dimensions of the inequivalent representations of $A$
does not exceed $\dim A$. This shows that each irreducible
representation of a finite-dimensional algebra is finite-dimensional
and there are only finitely many equivalence classes of such
representations. 

This is not such a strong statement as it seems because $A$ may have
indecomposable representations. In fact $A$ may have an infinite
number of inequivalent representations of a given dimension even if
$\dim A<\infty$. [{\it E.g} consider the three dimensional complex
algebra, consisting of 
$\lambda +\mu x + \nu y. \quad \lambda, \mu,\nu\in\Cop$, subject to   
$x^2=y^2=xy=yx=0$, which has the faithful three dimensional
representation 
$\pmatrix \lambda&0&\mu\cr 0&\lambda&\nu\cr 0&0&\lambda\endpmatrix$
and the inequivalent two-dimensional representations
$\pmatrix \lambda&\mu+\xi\nu\cr 0&\lambda\cr\endpmatrix$, for each 
$\xi\in\Cop$.] The situation is more under control if the algebra is
semi-simple. 

The ring $R$ is semi-simple if the adjoint representation is
completely reducible. If $R$ is semi-simple, $1$ is the sum of a
finite number of elements of $R$, one in each of a number of the
summands in the expression of $R$ as a sum of irreducible modules, 
$1=\sum_{i=1}^ne_i$ and, since any $r=\sum_{i=1}^nre_i$, it follows
that there is a finite number, $n$, of summands $R_i=Re_i$ and 
$R=\bigoplus_{i=1}^nR_i$. If $R$ is semi-simple, every $R$ module is
completely reducible (though not necessarily into a finite number of
irreducible summands). [Any module $M$ is the quotient of the free
module $\R=\bigoplus_{m\in M} R_m$, where $R_m$ is a copy of the
adjoint module $R$, by the ideal consisting of those $(r_m)_{m\in M}$
such that $\sum_{m\in M}r_mm=0$. The result follows since $\R$ is
completely reducible if $R$ is and the quotient of a completely
reducible module is itself completely reducible.] 

An $R$ module $M$ is finitely-generated if 
$M=\{\sum_{i=1}^n r_im_i:r_i\in R\}$ for a finite number, $n$, of
fixed elements $m_i\in M$, $1\leq i\leq n$. If $R$ is semi-simple, any
finitely generated $R$ module is completely reducible into a finite
number of summands. [This follows because $M$ is the quotient of
$\bigoplus_{i=1}^n R_i$,  where $R_i\cong R$, by the ideal
$\{(r_i):\sum_{i=1}^n r_im_i=0\}$.] 

If $M$ and $N$ are $R$ modules, an $R$-homomorphism $f:M\rightarrow N$
is a map satisfying $rf=fr$. If $M$ and $N$ are simple modules,
Schur's Lemma implies that the set of $R$-homomorphisms
$\Hom_R(M,N)=0$ if $M$ and $N$ are not equivalent. If $M=N$,
$\Hom_R(M,M)\equiv \End_R(M)$ is a division ring, that is every a ring
in which every non-zero element has an inverse. In the case of an
algebra, if $\dim M<\infty$, $\End_A(M)=F$, the underlying field. 

If the $R$-module $M$ is completely decomposable into a finite number
of irreducible submodules, we can write $M=\sum_{i=1}^N M^{n_i}_i$,
where each $M_i$ is irreducible and $M_i$ and $M_j$ are inequivalent
if $i\ne j$. Since $\Hom_R(M_i,M_j)=0$ if $i\ne j$, 
$$\End_R(M)=\prod_{i=1}^N\End_R(M_i^{n_i})=\prod_{i=1}^n
\M_{n_i}(D_i),$$
where the division algebra $D_i=\End_R(M_i)$ and $\M_n(D)$ 
is the ring of $n\times n$ matrices with entries in the division
algebra $D$.

If $R$ is a semi-simple ring, we can write
$R=\oplus_{i=1}^NR_i^{n_i}$, where the $R_i$ are irreducible as $R$
modules and inequivalent for $i\ne j$. So 
$\End_RR=\prod_{i=1}^N\M_{n_i}(D_i)$, where
$D_i=\End_R(R_i)$. But $\End_RR=R^o$, the reverse ring to $R$ defined
on the set $R$ by taking the product of $r$ and $s$ to be $sr$ rather
than $rs$. Since, evidently, $(R^o)^o=R$, 
$$R=\prod_{i=1}^N\M_{n_i}(D_i^o),$$ 
\ie\ every semi-simple ring is isomorphic to the direct product of a
finite number of finite-dimensional matrix rings over division
algebras [Wedderburn's Structure Theorem].

In the case of a semi-simple algebra, each $D_i=F$, the underlying
field, so 
$$A =\prod_{i=1}^N\M_{n_i}(F),$$
where $\M_n(F)$ is the algebra of $n\times n$ matrices with
entries in the field $F$. In particular, any semi-simple algebra is
finite-dimensional.

\vskip24pt
\leftline{\bf REFERENCES}
\vskip6pt
\item{[1]} G. Veneziano, {\it Construction of a crossing-symmetric,
Regge-behaved amplitude for linearly rising trajectories}, Nuovo
Cim. {\bf 57A}, 190 (1968).
\item{[2]} Z. Koba and H.B. Nielsen, {\it Manifestly crossing-invariant
parametrization of $n$-meson amplitude}, Nucl. Phys. {\bf B12}, 517
(1969). 
\item{[3]} A.A. Belavin, A.M. Polyakov and A.B. Zamolodchikov, 
{\it Infinite conformal symmetry in two-dimensional quantum field
theory}, Nucl. Phys. {\bf B241}, 333 (1984).
\item{[4]} G. Moore and N. Seiberg, {\it Classical and Quantum
Conformal Field Theory}, Comm. Math. Phys. {\bf 123}, 177 (1989). 
\item{[5]} G. Segal, {\it Notes on Conformal Field Theory},
unpublished manuscript. 
\item{[6]} R.E. Borcherds, {\it Vertex algebras, Kac-Moody algebras
and the monster},  Proc. Nat. Acad. Sci. U.S.A. {\bf 83}, 3068 (1986). 
\item{[7]} R.E. Borcherds, {\it Monstrous moonshine and monstrous Lie
algebras}, Invent. Math. {\bf 109}, 405 (1992).
\item{[8]} I. Frenkel, J. Lepowsky and A. Meurman, {\it Vertex Operator
Algebras and the Monster} (Academic Press, 1988).
\item{[9]} I. Frenkel, Y.-Z. Huang and J. Lepowsky, {\it On
axiomatic approaches to vertex operator algebras and modules},
Mem. Amer. Math. Soc. {\bf 104}, 1 (1993).
\item{[10]} V. Kac, {\it Vertex algebras for beginners}
(Amer. Math. Soc., 1997).
\item{[11]} A.J. Wassermann, {\it Operator algebras and 
conformal field theory} in {\it Proceedings of the 
I.C.M. Z\"urich 1994)} (Birkh\"auser, 1995) 966.
\item{[12]} F. Gabbiani and J. Fr\"ohlich, {\it Operator algebras and
conformal field theory}, Commun. Math. Phys. {\bf 155}, 569 (1993).
\item{[13]} P. Goddard, {\it Meromorphic conformal field theory} in 
{\it Infinite dimensional Lie algebras and Lie groups: Proceedings  
of the CIRM Luminy Conference, 1988} (World Scientific, Singapore,
1989) 556.
\item{[14]} K. Osterwalder and R. Schrader, {\it Axioms for euclidean
Green's functions}, Commun. Math. Phys. {\bf 31}, 83 (1973).
\item{[15]} G. Felder, J. Fr\"ohlich and G. Keller, {\it On the
structure of unitary conformal field theory. I. Existence of conformal
blocks}, Commun. Math. Phys. {\bf 124}, 417 (1989).
\item{[16]} Y.-Z. Huang, {\it A functional-analytic theory of vertex
(operator) algebras}, {\tt math.QA/9808022}.
\item{[17]} P. Montague, {\it  On Representations of Conformal Field  
Theories and the Construction of Orbifolds}, Lett. Math. Phys. 
{\bf 38}, 1 (1996), {\tt hep-th/9507083}.
\item{[18]} Y. Zhu, {\it Vertex Operator Algebras, Elliptic Functions 
and Modular Forms}, Caltech preprint (1990), J. Amer. Math. Soc. 
{\bf 9}, 237 (1996).
\item{[19} M.R. Gaberdiel and P. Goddard, in preparation.
\item{[20]} I.B. Frenkel and Y. Zhu, {\it Vertex operator algebras
associated to representations of affine and Virasoro algebras},
Duke Math. J. {\bf 66}, 123 (1992).
\item{[21]} C. Dong, H. Li, G. Mason, {\it Twisted Representations of 
Vertex Operator Algebras}, {\tt q-alg/9509005}.
\item{[22]} E. Hille, {\it Analytic Function Theory I} (Blaisdell
Publishing Company, 1959).
\item{[23]} E. Hille, {\it Analytic Function Theory II} (Blaisdell
Publishing Company, 1962).
\item{[24]} B. Farb, R.K. Dennis, {\it Noncommutative Algebra},
(Springer, 1993).
\bye